\documentclass[preprint,showpacs,preprintnumbers,amsmath,amssymb,showkeys]{revtex4}

\usepackage{graphicx}
\usepackage{dcolumn}
\usepackage{bm}

\begin{document}
 
\noindent

\preprint{}
 
\title{No-signaling non-unitary modification of quantum dynamics within a deterministic model of quantization} 
 
\author{Agung Budiyono}  
\email{agungbymlati@gmail.com}

\affiliation{Jalan Emas 772 Growong Lor RT 04 RW 02, Juwana, Pati, 59185 Jawa Tengah, Indonesia}

\date{\today}
 
\begin{abstract} 
We have developed in the previous works a statistical model of quantum fluctuation based on a chaotic deviation from infinitesimal stationary action which is constrained by the principle of Locality to have a unique exponential distribution up to a parameter that determines its average. The unitary Schr\"odinger time evolution with Born's statistical interpretation of wave function is recovered as a specific case when the average deviation from infinitesimal stationary action is given by $\hbar/2$ for all the time. This naturally suggests a possible generalization of the quantum dynamics and statistics by allowing the average deviation fluctuates effectively randomly around $\hbar/2$ with a finite yet very small width and a finite time scale. We shall show that averaging over such fluctuation will lead to a non-unitary average-energy-conserving time evolution providing an intrinsic mechanism of decoherence in energy basis in macroscopic regime. A possible cosmological origin of the fluctuation is suggested. Coherence and decoherence are thus explained as two features of the same statistical model corresponding to microscopic and macroscopic regimes, respectively. Moreover, noting that measurement-interaction can be treated in equal footing as the other types of interaction, the objective locality of the model is argued to imply no-signaling between a pair of arbitrarily separated experiments.         
\end{abstract}  
 
\maketitle  

\section{Motivation}    
 
There has long been considerable interest to search for possible generalizations of quantum mechanics with various motivations. One way to modify quantum mechanics is to give up its unitary time evolution for a closed system. {\it Non-unitary} modifications of quantum dynamics for example appear in various phenomenological models for objective wave function collapse \cite{Ghirardi 1,Ghirardi 2,Diosi,Gisin,Pearle,Ghirardi 3,Percival PSD,Penrose}. It also emerges from the models for `intrinsic' decoherence by considering a stochastic time \cite{Milburn,Garay,Gambini 1}, or by assuming possible gravitational effects due to the fluctuation of spacetime \cite{Ellis 1,Diosi decoherence,Gambini QG}. Another possible way to modify quantum mechanics is to assume that the linearity of its dynamical law for a closed system is {\it not} exact. A {\it nonlinear} modification of quantum dynamics therefore is sought to set precision tests against quantum mechanics. It has already been shown however that a nonlinear modification of quantum dynamics \cite{Weinberg nonlinearity} may lead to signaling  \cite{Gisin signaling,Polchinski signaling} thus is in a strict conflict with the theory of relativity.     
 
In the present paper we shall suggest a possible modification of the quantum dynamics which also leads effectively to a non-unitary time evolution in macroscopic regime. Unlike the above mentioned various ad-hoc modifications, however, we shall start by reconstructing quantum mechanics from a statistical model. This is done by developing a deterministic model for the apparent random behavior of the microscopic world based on a chaotic deviation from infinitesimal stationary action reported in Refs. \cite{AgungSMQ6,AgungSMQ4,AgungSMQ7,AgungSMQ8,AgungSMQ9}. We have shown in Refs. \cite{AgungSMQ6,AgungSMQ8} that imposing the principle of Locality of no-influence-at-a-distance singles out uniquely an exponential distribution of deviation from infinitesimal action, up to a free parameter $\lambda$ which determines its average deviation as $|\lambda/2|$. We have also shown that such a statistical model leads to a `generalized Schr\"odinger equation' in which the reduced Planck constant $\hbar$ of the usual Schr\"odinger equation is replaced by $|\lambda|$ (see Eq. (\ref{generalized Schroedinger equation particle in potentials}) below). Hence the unitary quantum dynamics is regained as a specific case of the statistical model when $|\lambda|$ is constant for all time given by $\hbar$, so that the statistical model is stationary in time and the average deviation from infinitesimal stationary action is given by $\hbar/2$. 

It is thus instructive to go beyond the stationary case by allowing the free parameter $|\lambda|$ fluctuates chaotically with time around $\hbar$ with a small finite width and a finite time scale. In such a case, we have a randomly parameterized time evolution, giving the unitary Schr\"odinger evolution as zeroth order approximation. We shall show that averaging over the fluctuation of $|\lambda|$ around $\hbar$ will lead to a non-unitary time evolution, providing an intrinsic mechanism of decoherence in energy basis, adding to the environment-induced-decoherence \cite{Zeh 1,Zurek 1,Joos book,Schlosshauer,Nieuwenhuizen}. The rate of the loss of coherence is determined by the time scale, average and mean deviation of the fluctuation of $|\lambda|$. Coherence and decoherence are thus explained as two features of the same statistical model corresponding to microscopic and macroscopic regimes, respectively. Within the non-unitary modified dynamics, the quantum mechanically conserved quantity, including the average energy, is kept conserved, and von Neumann entropy is monotonically increasing.    

Since $|\lambda|$ is left unfixed by the principle of Locality, it must be a global parameter uniform across the space. This suggests that the statistical fluctuation of $|\lambda|$, which leads both to quantum coherence and its loss (decoherence) in macroscopic regime, may have a cosmological origin. Moreover, since the model respects the principle of Locality objectively (independence of measurement), and measurement-interaction can be treated in equal footing as the other types of interaction \cite{AgungSMQ7}, we shall argue that the objective locality of the model implies no-signaling in the context of measurement \cite{AgungSMQ9}. We have thus a class of no-signaling modification of quantum mechanics which leads to an intrinsic non-dissipative mechanism of decoherence.    

\section{A deterministic statistical model of the universal microscopic randomness}  
 
\subsection{A class of statistical model of microscopic randomness based on a chaotic fluctuation of infinitesimal stationary action}

It is remarkable that, hitherto, despite of the continuous pragmatical successes shown by quantum mechanics, there is no consensus on the nature and origin of the randomness of microscopic physical phenomena. Standard quantum mechanics {\it does} provide a set of `abstract rules to calculate' with unparalleled accuracy the statistical results of measurement in an ensemble of identically prepared experiment, yet it says almost nothing on the most tantalizing question: {\it what fluctuates where and why?} \cite{Pearle why}. One is therefore kept wondering to ask along with Wheeler \cite{Wheeler}: {\it why the rules?} Moreover, the prediction of quantum mechanics on the AB (Aharonov-Bohm) effect \cite{Aharonov-Bohm} and its experimental verification \cite{Peshkin} suggest that the randomness in microscopic regime is inexplicable in term of conventional random forces as in the Brownian motion, unless a sort of nonlocality is allowed \cite{Aharonov non-local AB effect}. 
  
To investigate the above foundational question, let us first discuss a statistical model of the apparently random microscopic deviations from classical mechanics based on a deterministic chaotic fluctuation of infinitesimal stationary action reported in Refs. \cite{AgungSMQ6,AgungSMQ4,AgungSMQ7,AgungSMQ8,AgungSMQ9}. Let $q$ denotes the configuration of the system and $t$ is time parameterizing the evolution of the system. Let us assume that the Lagrangian depends on a `randomly fluctuating variable $\xi(t)$': $L=L(q,\dot{q};\xi(t))$, whose physical origin will be suggested later. Note that by `a randomly fluctuating $\xi(t)$', we mean that $\xi$ is a {\it deterministic}, smooth and differentiable function of $t$, yet its behavior is sufficiently {\it chaotic} so that it can be regarded as {\it effectively random}. Let us assume that the time scale for the fluctuation of $\xi$ is $dt$, so that during a time interval of length $dt$, $\xi$ is ``effectively'' constant (its variation is negligible). Let us then consider two infinitesimally close spacetime points $(q;t)$ and $(q+dq;t+dt)$ such that $\xi$ is effectively constant. Let us assume that fixing $\xi$, the principle of stationary action is valid to select a segment of path, denoted by $\mathcal{J}(\xi)$, that connects the two points. One must then solve a variational problem with fixed end points: $\delta(Ldt)=0$. This leads to the existence of the Hamilton's principal function $A(q;t,\xi)$ whose differential along the path is given by \cite{Rund book}, for a fixed $\xi$,  
\begin{equation} 
dA=Ldt=p\cdot dq-Hdt,  
\label{infinitesimal stationary action} 
\end{equation} 
where $p(\dot{q})=\partial L/\partial{\dot{q}}$ is the momentum and $H(q,p;\xi)\doteq p\cdot\dot{q}(p)-L(q,\dot{q}(p);\xi)$ is the Hamiltonian which is also parameterized by $\xi(t)$. The above relation implies the following Hamilton-Jacobi equation:
\begin{eqnarray}
p=\partial_qA,\hspace{10mm}\nonumber\\
-H(q,p;\xi)=\partial_tA, 
\label{Hamilton-Jacobi condition}
\end{eqnarray}
now parameterized by a chaotically fluctuating $\xi(t)$. Hence, $dA(\xi)$ is just the `infinitesimal stationary action' along the corresponding short path during the infinitesimal time interval $dt$ with a constant $\xi$. 

Varying $\xi$, the principle of stationary action will therefore pick up various different paths $\mathcal{J}(\xi)$, all connecting the same two infinitesimally close spacetime points in configuration space, each with possibly different values of infinitesimal stationary action $dA(\xi)$. Since $\xi(t)$ is sufficiently chaotic, $dA(\xi)$ is thus {\it effectively} random, and so are the momentum, velocity and energy of the system. A single event description is thus practically impossible, a statistical approach is necessary. Hence, we have a dynamical processes in which the system starting with a configuration $q$ at time $t$ may take various different paths in configuration space effectively randomly to end up with a configuration $q+dq$ at time $t+dt$. The dynamical processes is thus completely determined by a `transition probability' for the system starting with a configuration $q$ at time $t$ to move to its infinitesimally close neighbor $q+dq$ at time $t+dt$ via a path $\mathcal{J}(\xi)$, below denoted by 
\begin{equation}
P((q+dq;t+dt)|\{\mathcal{J}(\xi),(q;t)\}).  
\label{transition probability}
\end{equation}

It is natural to assume that the transition probability depends on the chaotic quantity $dA(\xi)$ evaluated along the short segment of trajectory. Further, in view of the micro-macro correspondence, then it is reasonable to assume that the transition probability is a function of a quantity that measures the deviation from classical mechanics. Let us first assume that $\xi$ is the simplest random variable with two possible values, a binary random variable. Without losing generality let us assume that the two possible values of $\xi$ differ from each other only by their signs, namely one is the opposite of the other, $\xi=\pm|\xi|$. Let us suppose that both realizations of $\xi$ lead to the same path so that $dA(\xi)=dA(-\xi)$. Since the stationary action principle is valid for both values of $\pm\xi$, then such a binary model must recover classical mechanics. Hence, in this simple model, the non-classical behavior must correspond to the case when $dA(\xi)\neq dA(-\xi)$.  

Now let us proceed to assume that $\xi$ may take continuous values. Let us assume that even in this case the difference of the values of $dA$ at $\pm\xi$, 
\begin{equation}
Z(q;t,\xi)\doteq dA(q;t,\xi)-dA(q;t,-\xi)=-Z(q;t,-\xi),  
\label{the fluctuation}
\end{equation}
measures the non-classical behavior of the stochastic process, namely the larger the difference, the stronger is the deviation from classical mechanics. Hence $Z(\xi)$ is randomly fluctuating due to the chaotic fluctuation of $\xi$, and we shall use the distribution of its magnitude as the transition probability to construct the stochastic model: 
\begin{equation}
P((q+dq;t+dt)|\{\mathcal{J}(\xi),(q;t)\})=P(Z(\xi)). 
\end{equation}

One can see that the randomness is built into the statistical model in a fundamentally different way from that of the classical Brownian motion. First, unlike the latter which is inherently random, the statistical model is in principle deterministic: the randomness is only apparent, due to the chaotic fluctuation of $\xi(t)$. Moreover, unlike the Brownian motion in which the randomness is introduced by adding some random forces, the model is based on a random fluctuation of infinitesimal stationary action due to the chaotic fluctuation of the variable $\xi(t)$ parameterizing the Lagrangian $L(q,\dot{q};\xi)$. One may expect that this will explain the physical origin of AB effect without invoking any kind of influence-at-a-distance or nonlocality. 

As another difference, recall that in the Newtonian formalism for a stochastic dynamics of interacting many particles compound system, it is always possible to introduce a set of random forces acting locally to each single particle. Accordingly, one can always define a joint-probability distribution of the random forces so that the randomness of the whole compound system can be said to originate from the combination of the randomness of each subsystem and their correlation, the usual bottom-up information flow. By contrast, in the statistical model, since the origin of the randomness is due to the chaotic fluctuation of the infinitesimal stationary action, and action is evaluated in configuration space instead of in ordinary space, then for interacting many particles system, the randomness of the whole compound system cannot be regarded as arising from the randomness of each subsystem. Rather it is the other way around: the randomness of each subsystem is induced by the randomness of the whole compound system, a manifestation of the top-down causation \cite{Ellis}, or the randomness of the whole gives a global constraint/context to the randomness of each subsystem. In this sense, there is a {\it statistical inseparability} of the randomness in the whole interacting compound system.   

\subsection{Exponential distribution of infinitesimal stationary action as the transition probability}
   
To proceed, one must therefore fix the distribution of $Z(\xi)$ defined in Eq. (\ref{the fluctuation}). First, it is reasonable to assume that the transition probability must be decreasing as the non-classicality becomes stronger. Hence, the transition probability must be a decreasing function of the absolute value of $Z(\xi)$. There are however infinitely many such probability distribution functions. The question is then: what selects one of them to describe the microscopic world? To this end, we have argued in Refs. \cite{AgungSMQ6,AgungSMQ8} that imposing the principle of Locality and Macroscopic Classicality leads uniquely to a class of probability distribution given by the following exponential law:
\begin{eqnarray}
P(Z)\propto N e^{\frac{1}{\lambda(\xi)}Z(\xi)}=N e^{\frac{1}{\lambda(\xi)}\big(dA(\xi)-dA(-\xi)\big)},
\label{exponential law 1st}
\end{eqnarray}
where $N$ is a factor independent of $Z(\xi)$ whose form to be specified later, and $\lambda(\xi)$ is a {\it non-vanishing} function of $\xi$ with action dimensional, thus is in general randomly fluctuating with time. As will be discussed later, $\lambda$ must be independent of the configuration of the system.  

Notice that by definition given in Eq. (\ref{the fluctuation}), $Z(\xi)$ changes its sign as $\xi$ flips its sign: $Z(-\xi)=-Z(\xi)$. On the other hand, to guarantee the negative definiteness of the exponent in Eq. (\ref{exponential law 1st}) for normalizability at any spacetime point, $\lambda$ must always have the opposite sign of $Z(\xi)$. This demands that $\lambda$ must flip its sign as $\xi$ changes its sign. This fact allows us to assume that both $\lambda(\xi)$ and $\xi$ always have the {\it same} sign. The time scale for the fluctuation of the sign of $\lambda$ is therefore the same as that of $\xi$ given by $dt$. 

It is however clear that for the distribution of Eq. (\ref{exponential law 1st}) to make sense mathematically, the time scale for the fluctuation of $|\lambda|$ must be much larger than the time scale of the fluctuation of the magnitude of the envelope of $\xi$, below denoted by $\|\xi\|$. Let us then denote them respectively as $\tau_{\lambda}$ and $\tau_{\xi}$. Let us further assume that $\tau_{\xi}$ is much larger than $dt$. One thus has 
\begin{equation}
\tau_{\lambda}\gg\tau_{\xi}\gg dt. 
\label{time scales}
\end{equation}
Hence, in a time interval of length $\tau_{\xi}$, the magnitude of the envelope of $\xi$ is effectively constant while the sign of $\xi$ may fluctuate randomly together with the sign of $\lambda$ in a time scale $dt$. Moreover, in a time interval of length $\tau_{\lambda}$, $|\lambda|$ is effectively constant and $\|\xi\|$ fluctuates randomly so that the distribution of $|Z(\xi)|$ is given by the exponential law of Eq. (\ref{exponential law 1st}) characterized by $|\lambda|$.    

Next, let us introduce a new chaotic quantity $S(q;t,\xi)$ so that the differential along the path $\mathcal{J}(\xi)$ is given by 
\begin{equation}
dS(q;t,\xi)=\frac{dA(q;t,\xi)+dA(q;t,-\xi)}{2}=dS(q;t,-\xi).  
\label{infinitesimal symmetry}
\end{equation}
Subtracting $dA(q;t,\xi)$ from both sides, one has 
\begin{equation}
dS(q;t,\xi)-dA(q;t,\xi)=\frac{dA(q;t,-\xi)-dA(q;t,\xi)}{2}.
\label{average of classical deviation}
\end{equation}  
Using $dS$, the transition probability of Eq. (\ref{exponential law 1st}) can thus be written as 
\begin{eqnarray} 
P((q+dq;t+dt)|\{\mathcal{J}(\xi),(q;t)\})\hspace{15mm}\nonumber\\
\propto N e^{-\frac{2}{\lambda}(dS(q;t,\xi)-dA(q;t,\xi))}\doteq P_S(dS|dA).  
\label{exponential distribution of DISA}
\end{eqnarray}  

Since $dA(\xi)$ is just the infinitesimal stationary action along the path $\mathcal{J}(\xi)$, then we shall refer to $dS(\xi)-dA(\xi)$ as the deviation from infinitesimal stationary action. One may therefore see the above transition probability to be given by an exponential distribution of deviation from infinitesimal stationary action $dS-dA$ parameterized by $|\lambda|$. It can also be regarded as the conditional probability density of $dS$ given $dA$, suggesting the use of the notation $P_S(dS|dA)$. The relevancy of such an exponential law to model the apparently stochastic, microscopic deviation from classical mechanics is firstly suggested in Ref. \cite{AgungSMQ4}. An application of the deterministic model to reconstruct quantum measurement is reported in Refs. \cite{AgungSMQ7,AgungSMQ9}.

Further, there is no a priori reason on how the sign of $dS(\xi)-dA(\xi)$, which is equal to the sign of $Z(-\xi)=dA(-\xi)-dA(\xi)$ due to Eq. (\ref{average of classical deviation}), should be distributed at any given spacetime point. The principle of insufficient reason (principle of indifference) \cite{Jaynes book} then suggests to assume that at any spacetime point, there is equal probability for $dS-dA$ to take positive or negative values. Since the sign of $dS(\xi)-dA(\xi)$ changes as $\xi$ flips its sign, then the sign of $\xi$ must also be distributed equally probably. Hence, the probability density of the occurrence of $\xi$ at any time, denoted below by $P_{H}(\xi)$, must satisfy the following unbiased condition:  
\begin{equation}
P_{H}(\xi)=P_{H}(-\xi).  
\label{God's unbiased}
\end{equation}  
Let us note that $P_{H}(\xi)$ may depend on time, thus it is in general not stationary. Since the sign of $\lambda$ is always the same as that of $\xi$, then the probability of the occurrence of $\lambda$ must also satisfy the same unbiased condition $P(\lambda)=P(-\lambda)$.  

Fixing $|\lambda|$ in Eq. (\ref{exponential distribution of DISA}) which is valid for a time interval less then $\tau_{\lambda}$, then the average deviation from infinitesimal stationary action is given by
\begin{equation}
\overline{|dS-dA|}=|\lambda|/2. 
\label{the P} 
\end{equation}
Note however that since in general $|\lambda|$ may fluctuate with time, then in general $P_S(dS|dA)$ is {\it not} stationary. It is stationary only when $|\lambda|$ is constant for all the time, that is when $\tau_{\lambda}=\infty$. We shall argue later that the stationary case leads to quantum coherence, while the non-stationary case ($\tau_{\lambda}$ is finite) leads to loss of coherence (decoherence) in macroscopic regime.  

One can also see that in the regime where the average deviation from infinitesimal stationary action is much smaller than the infinitesimal stationary action itself, namely $|dA|/|\lambda|\gg 1$, or formally in the limit $|\lambda|\rightarrow 0$, Eq. (\ref{exponential distribution of DISA}) reduces to  
\begin{equation}
P_S(dS|dA)\rightarrow \delta(dS-dA), 
\label{macroscopic classicality}
\end{equation}
or $dS(\xi)\rightarrow dA(\xi)$, so that $S$ satisfies the Hamilton-Jacobi equation of (\ref{Hamilton-Jacobi condition}) by virtue of Eq. (\ref{infinitesimal stationary action}). Due to Eq. (\ref{average of classical deviation}), in this regime one has $|dA(\xi)-dA(-\xi)|\rightarrow 0$. Hence such a limiting case must be identified to correspond to the macroscopic regime. This observation suggests that $|\lambda|$ must take microscopic values. Moreover, in this limit one also has $Z(\xi)=dA(\xi)-dA(-\xi)\rightarrow 0$. Noting that $Z(\xi)=-Z(-\xi)$, one may thus conclude that the limit is realized when $|\xi|\rightarrow 0$. That is, in this regime, the Lagrangian is effectively independent of $\xi$ given by the usual classical Lagrangian.    

\subsection{Modified Hamilton-Jacobi equation} 
  
Let us now derive a set of differential equations which characterizes the dynamical processes when the transition probability is given by Eq. (\ref{exponential distribution of DISA}) valid for a microscopic time scale $\tau_{\xi}$ \cite{AgungSMQ6,AgungSMQ4,AgungSMQ7,AgungSMQ8}. Let us consider a time interval of length $\tau_{\lambda}$ in which $|\lambda|$ is effectively constant. Recall that since $\tau_{\lambda}\gg\tau_{\xi}\gg dt$, then within this time interval, $|dS(\xi)-dA(\xi)|$ fluctuates randomly due to the fluctuation of $\|\xi\|$, distributed according to the exponential law of Eq. (\ref{exponential distribution of DISA}) characterized by $|\lambda|$. 

Let us then denote the joint-probability density that at time $t$ the configuration of the system is $q$ and a random value of $\xi$ is realized by $\Omega(q,\xi;t)$. The marginal probability densities thus read 
\begin{eqnarray}
\rho(q;t)\doteq\int d\xi\Omega(q,\xi;t),\hspace{2mm}P_{H}(\xi)=\int dq \Omega(q,\xi;t).\hspace{0mm}
\label{marginal probabilities general}
\end{eqnarray} 
To comply with Eq. (\ref{God's unbiased}), the joint-probability density must satisfy the following symmetry relation: 
\begin{eqnarray} 
\Omega(q,\xi;t)=\Omega(q,-\xi;t). 
\label{God's fairness}
\end{eqnarray} 
From Eq. (\ref{infinitesimal symmetry}), one also has, fixing $\xi$, the following symmetry relations: 
\begin{eqnarray}
\partial_qS(q;t,\xi)=\partial_qS(q;t,-\xi),\nonumber\\
\partial_tS(q;t,\xi)=\partial_tS(q;t,-\xi). 
\label{quantum phase symmetry}
\end{eqnarray} 
Both Eqs. (\ref{God's fairness}) and (\ref{quantum phase symmetry}) will play important roles later.  

Let us then evolve $\Omega(q,\xi;t)$ along a time interval $\Delta t$ with $\tau_{\xi}\ge\Delta t\gg dt$ so that $\|\xi\|$ is effectively constant while the sign of $\xi$ fluctuates randomly. Given a constant $\xi$, let us consider two infinitesimally close spacetime points $(q;t)$ and $(q+dq;t+dt)$. Let us assume that for this value of $\xi$, the two points are connected to each other by a segment of trajectory $\mathcal{J}(\xi)$ picked up by the principle of stationary action so that the differential of $S(\xi)$ along this segment is $dS(\xi)$, parameterized by $\xi$. Then according to the conventional probability theory, the conditional joint-probability density that the system initially at $(q;t)$ traces the segment of trajectory $\mathcal{J}(\xi)$ and end up at $(q+dq;t+dt)$, denoted below as $\Omega\big(\{(q+dq,\xi;t+dt),(q,\xi;t)\}\big|\mathcal{J}(\xi)\big)$, is equal to the probability that the configuration of the system is $q$ at time $t$, $\Omega(q,\xi;t)$, multiplied by the transition probability between the two infinitesimally close points via the segment of trajectory $\mathcal{J}(\xi)$ which is given by Eq. (\ref{exponential distribution of DISA}). One thus has  
\begin{eqnarray}
\Omega\Big(\{(q+dq,\xi;t+dt),(q,\xi;t)\}\big|\mathcal{J}(\xi)\Big)\hspace{15mm}\nonumber\\
=P((q+dq;t+dt)|\{\mathcal{J}(\xi),(q;t)\})\times\Omega(q,\xi;t)\nonumber\\
\propto Ne^{-\frac{2}{\lambda}(dS(\xi)-dA(\xi))}\times\Omega(q,\xi;t).  
\label{probability density} 
\end{eqnarray}    

The above equation describing the dynamics of ensemble of trajectories must give back the time evolution of classical mechanical ensemble of trajectories when $S$ approaches $A$. This requirement puts a constraint on the functional form of the factor $N$ in Eq. (\ref{exponential distribution of DISA}). To see this, let us assume that $N$ takes the following general form: 
\begin{equation}
N\propto\exp(-\theta(S)dt), 
\label{exponential classical}
\end{equation}
where $\theta$ is a scalar function of $S$. Inserting this into Eq. (\ref{probability density}), taking the limit $S\rightarrow A$ and expanding the exponential up to the first order one gets $\Omega\big(\{(q+dq,\xi;t+dt),(q,\xi;t)\}\big|\mathcal{J}(\xi)\big)\approx \big[1-\theta(A)dt\big]\Omega(q,\xi;t)$, which can be further written as $d\Omega=-\big(\theta(A)dt\big)\Omega$, 
where $d\Omega(q,\xi;t)\doteq\Omega\big(\{(q+dq,\xi;t+dt),(q,\xi;t)\}\big|\mathcal{J}(\xi)\big)-\Omega(q,\xi;t)$ is the change of the probability density $\Omega$ due to the transport along the segment of trajectory $\mathcal{J}(\xi)$. Dividing both sides by $dt$ and taking the limit $dt\rightarrow 0$, one obtains $\dot{\Omega}+\theta(A)\Omega=0$. To guarantee a smooth correspondence with classical mechanics, the above equation must be identified as the continuity equation describing the dynamics of ensemble of classical trajectories. To do this, it is sufficient to choose $\theta(S)$ to be determined uniquely by the classical Hamiltonian as \cite{AgungSMQ6,AgungSMQ4,AgungSMQ7,AgungSMQ8}  
\begin{equation}
\theta(S)=\partial_q\cdot\Big(\frac{\partial H}{\partial p}\Big|_{p=\partial_qS}\Big), 
\label{QC correspondence}
\end{equation}
so that in the limit $S\rightarrow A$, it is given by the divergence of the corresponding classical velocity field.  

Now let us consider the case when $|(dS-dA)/\lambda|\ll 1$. Again, inserting Eq. (\ref{exponential classical}) into Eq. (\ref{probability density}) and expanding the exponential on the right hand side up to the first order one gets
\begin{eqnarray}
d\Omega=-\Big[\frac{2}{\lambda}(d S-dA)+\theta(S)d t\Big]\Omega.    
\label{fundamental equation 0}
\end{eqnarray} 
Further, recalling that $\xi$ is effectively constant during the infinitesimal time interval $dt$, one can expand the differentials $d\Omega$ and $dS$ in Eq. (\ref{fundamental equation 0}) as $dF=\partial_tF dt+\partial_qF\cdot dq$. Using Eq. (\ref{infinitesimal stationary action}), one finally obtains the following pair of coupled differential equations:  
\begin{eqnarray}
p(\dot{q})=\partial_qS+\frac{\lambda}{2}\frac{\partial_q\Omega}{\Omega},\hspace{15mm}\nonumber\\
-H(q,p(\dot{q});\xi)=\partial_tS+\frac{\lambda}{2}\frac{\partial_t\Omega}{\Omega}+\frac{\lambda}{2}\theta(S). 
\label{fundamental equation rederived}
\end{eqnarray} 

Some notes are in order. First, the above pair of relations are valid when $\xi$ is fixed. However, since as discussed above $P_S(dS|dA)$ is insensitive to the sign of $\xi$ which is always equal to the sign of $\lambda$, then the above pair of equations are valid in a microscopic time interval of length $\tau_{\xi}$ during which $\|\xi\|$, and also $|\lambda|$ due to Eq. (\ref{time scales}), are constant, while the signs of $\xi$ and $\lambda$ change randomly. To have an evolution for a finite time interval $\tau_{\lambda}>\Delta t>\tau_{\xi}$, one may proceed along the following approximation. First, one divides the time into a series of intervals of length $\tau_{\xi}$: $t\in[(k-1)\tau_{\xi},k\tau_{\xi})$, $k=1,2,\dots$, and attributes to each interval a random value of $\xi(t)=\xi_k$ according to the probability distribution $P_{H_k}(\xi_k)=P_{H_k}(-\xi_k)$. Hence, during the interval $[(k-1)\tau_{\xi},k\tau_{\xi})$, the magnitude of $\xi=\xi_k$ is constant while its sign changes randomly in an infinitesimal time scale $dt$, so that Eq. (\ref{fundamental equation rederived}) is valid. One then applies the pair of equations in (\ref{fundamental equation rederived}) during each interval of time with a constant $|\xi(t)|=|\xi_k|$, consecutively. Moreover, to have a time evolution for $\Delta t\ge \tau_{\lambda}$, which is the main interest of the present paper, one must take into account the fluctuation of $|\lambda|$ with time.      

It is evident that as expected, in the formal limit $|\lambda|\rightarrow 0$, Eq. (\ref{fundamental equation rederived}) reduces back to the Hamilton-Jacobi equation of (\ref{Hamilton-Jacobi condition}). In this sense, Eq. (\ref{fundamental equation rederived}) can be regarded as a generalization of the Hamilton-Jacobi equation due to the chaotic deviation from infinitesimal stationary action following the exponential law of Eq. (\ref{exponential distribution of DISA}). Unlike the Hamilton-Jacobi equation in which we have a single unknown function $A$, however, to calculate the velocity or momentum and energy, one now needs a pair of unknown functions $S$ and $\Omega$. The pair of relations in Eq. (\ref{fundamental equation rederived}) must {\it not} be interpreted that the momentum and energy of the particles are determined causally by the gradient of the probability density $\Omega$ (or $\ln(\Omega)$), which is physically absurd. Rather it is the other way around as shown explicitly by Eq. (\ref{fundamental equation 0}). The relation is thus kinematical rather than causal-dynamical.          

\subsection{The principle of Locality and global variable\label{global lambda}}
 
Let us proceed to show that the above statistical model obeys the principle of Locality demanded by the theory of relativity that due to the finite maximum velocity of interaction, two subsystems spacelike separated from each other can {\it not} influence each other \cite{AgungSMQ6,AgungSMQ8}. To see this, it is sufficient to consider a compound system composed of two particles whose configuration is denoted by $q=(q_1,q_2)$, sufficiently remotely separated from each other so that due to the principle of Locality, there is no physical interaction connecting the two particles. The Lagrangian is thus decomposable as $L(q_1,q_2,\dot{q}_1,\dot{q}_2;\xi)=L_1(q_1,\dot{q}_1;\xi)+L_2(q_2,\dot{q}_2;\xi)$, so that the infinitesimal stationary action is also decomposable: $dA(q_1,q_2;\xi)=dA_1(q_1;\xi)+dA_2(q_2;\xi)$, and accordingly one has $dS(q_1,q_2;\xi)=dS_1(q_1;\xi)+dS_2(q_2;\xi)$ by virtue of Eq. (\ref{infinitesimal symmetry}). Here we have assumed that $\xi$ is a global variable. On the other hand, since the classical Hamilton $H$ is decomposable as $H(q_1,q_2,p_1,p_2;\xi)=H_1(q_1,p_1;\xi)+H_2(q_2,p_2;\xi)$, $p_{i}$, $i=1,2$, is the classical momentum of the $i-$particle, then $\theta$ of Eq. (\ref{QC correspondence}) is also decomposable: $\theta(q_1,q_2;\xi)=\theta_1(q_1;\xi)+\theta_2(q_2;\xi)$. 

Inserting all these into Eqs. (\ref{exponential classical}) and (\ref{exponential distribution of DISA}), then one can see that the distribution of deviation from infinitesimal stationary action for the two non-interacting remotely separated particles is separable as 
\begin{equation}
P_S(dS_1+dS_2|dA_1+dA_2)=P_S(dS_1|dA_1)P_S(dS_2|dA_2).   
\label{principle of Local Causality}
\end{equation}
Namely the joint-probability distribution of the deviation from infinitesimal stationary action of the compound of two particles is separable into the probability distribution of the deviation with respect to each single particle (subsystem). They are thus independent of each other, as intuitively expected for non-interacting remotely separated particles. It is therefore evident that the dynamics and statistics of the first particle has {\it no} influence whatsoever on the dynamics and statistics of the second particle and vice versa.         

Note that the statistical separability for non-interacting particles described by Eq. (\ref{principle of Local Causality}) is {\it unique} to the exponential law. A Gaussian distribution of deviation from infinitesimal stationary action for example does {\it not} have such a property. Hence, for a Gaussian distribution, a pair of spacelike separated non-interacting particles may still influence each other, which therefore contradicts the theory of relativity. Noting the above fact, we have argued in Refs. \cite{AgungSMQ6,AgungSMQ8} that the exponential law of the transition probability of Eq. (\ref{exponential distribution of DISA}) is the {\it unique} form of probability distribution of the deviation from infinitesimal stationary action, up to the free parameter $\lambda$ which determines its average deviation as $|\lambda|/2$, that is {\it singled out} among the infinitude of possibilities by the principle of Locality and Macroscopic Classicality, mathematically represented respectively by Eqs. (\ref{principle of Local Causality}) and (\ref{macroscopic classicality}). 

Finally, since $|\lambda|$ is left unfixed by the principle of Locality, then it must be independent of any local variables. In particular, it must be independent of the configuration of the system. Hence, one may conclude that $|\lambda|$ should be a global variable which depends only on the global variable $\xi$. Namely, at any given (cosmic) time, the value of $|\lambda|$ must be spatially uniform across the universe. This naturally suggests that the fluctuation of $|\lambda|$, and thus also $\xi$ or the infinitesimal stationary action, might be cosmological in origin (see below).  
    
\section{Quantization}  

\subsection{The Schr\"odinger equation and Born's statistical interpretation} 

Let us apply the above general formalism to statistically modify a concrete classical mechanical system. Let us first consider a system of a single particle subjected to external potentials so that the classical Hamiltonian takes the following general form:
\begin{equation}
H(q,p;\xi)=\frac{g^{ij}(q;\xi)}{2}(p_i-a_i)(p_j-a_j)+V,   
\label{classical Hamiltonian}
\end{equation} 
where $a_i(q)$, $i=x,y,z$ and $V(q)$ are vector and scalar potentials respectively, the metric $g^{ij}(q;\xi)$ may in general depend on the position of the particle, and summation over repeated indices are assumed. We have also assumed that the Hamiltonian is parameterized by $\xi$ via the metric. That is we have a chaotically fluctuating metric. The calculations in the present subsection have been reported in Refs. \cite{AgungSMQ6,AgungSMQ4,AgungSMQ7,AgungSMQ8}. Here we shall reproduce it as a reference for later discussion in the subsequence sections. Application to many particles system with different kind of classical Hamiltonians can be done in the same way by following exactly all the steps that we shall take below.      
 
Let us first consider a time interval of length $\tau_{\lambda}$ during which the absolute value of  $\lambda$ is effectively constant while its sign fluctuates randomly together with the random fluctuation of the sign of $\xi$ in a time scale $dt$. Moreover, let us then consider a microscopic time interval of length $\tau_{\xi}$ so that the magnitude of the envelope of $\xi$ is effectively constant while the sign of $\xi$ fluctuates randomly with equal probability. Within this interval of time, using Eq. (\ref{classical Hamiltonian}) to express $\dot{q}$ in term of $p$ via the (kinematic part of the) Hamilton equation $\dot{q}=\partial H/\partial p$, one has, from the upper equation of (\ref{fundamental equation rederived}) 
\begin{equation}
\dot{q}^i(\xi)=g^{ij}(\xi)\Big(\partial_{q_j}S(\xi)+\frac{\lambda}{2}\frac{\partial_{q_j}\Omega(\xi)}{\Omega(\xi)}-a_j\Big). 
\label{classical velocity field HPF particle in potentials}
\end{equation}
Assuming the conservation of probability one thus obtains the following continuity equation which gives a constraint to the dynamics of the ensemble:
\begin{eqnarray}
0=\partial_t\Omega+\partial_q\cdot(\dot{q}\Omega)\hspace{45mm}\nonumber\\
=\partial_t\Omega+\partial_{q_i}\Big(g^{ij}(\partial_{q_j}S-a_j)\Omega\Big)+\frac{\lambda}{2}\partial_{q_i}(g^{ij}\partial_{q_j}\Omega). 
\label{FPE particle in potentials}
\end{eqnarray} 

On the other hand, from Eq. (\ref{classical Hamiltonian}), $\theta(S)$ of Eq. (\ref{QC correspondence}) reads
\begin{equation}
\theta(S)=\partial_{q_i}g^{ij}(\partial_{q_j}S-a_j). 
\end{equation}
Using the above form  of $\theta(S)$, the lower equation of (\ref{fundamental equation rederived}) thus becomes
\begin{eqnarray}
-H(q,p;\xi)=\partial_tS+\frac{\lambda}{2}\frac{\partial_t\Omega}{\Omega}+\frac{\lambda}{2}\partial_{q_i}g^{ij}(\partial_{q_j}S-a_j).
\label{fundamental equation particle in potentials} 
\end{eqnarray}
Plugging the upper equation of (\ref{fundamental equation rederived}) into the left hand side of Eq. (\ref{fundamental equation particle in potentials}) and using Eq. (\ref{classical Hamiltonian}) one has, after arrangement 
\begin{eqnarray}
\partial_tS+\frac{g^{ij}}{2}(\partial_{q_i}S-a_i)(\partial_{q_j}S-a_j)+V\hspace{30mm}\nonumber\\
-\frac{\lambda^2}{2}\Big(g^{ij}\frac{\partial_{q_i}\partial_{q_j}R}{R}+\partial_{q_i}g^{ij}\frac{\partial_{q_j}R}{R}\Big)\hspace{30mm}\nonumber\\
+\frac{\lambda}{2\Omega}\Big(\partial_t\Omega+\partial_{q_i}\Big(g^{ij}(\partial_{q_j}S-a_j)\Omega\Big)+\frac{\lambda}{2}\partial_{q_i}(g^{ij}\partial_{q_j}\Omega)\Big)=0,
\label{HJM particle in potentials 0}
\end{eqnarray}
where we have defined $R\doteq\sqrt{\Omega}$ and used the identity:
\begin{equation}
\frac{1}{4}\frac{\partial_{q_i}\Omega}{\Omega}\frac{\partial_{q_j}\Omega}{\Omega}=\frac{1}{2}\frac{\partial_{q_i}\partial_{q_j}\Omega}{\Omega}-\frac{\partial_{q_i}\partial_{q_j}R}{R}. 
\label{fluctuation decomposition}
\end{equation}
Inserting Eq. (\ref{FPE particle in potentials}), the last line of Eq. (\ref{HJM particle in potentials 0}) vanishes to give
\begin{eqnarray}
\partial_tS+\frac{g^{ij}}{2}(\partial_{q_i}S-a_i)(\partial_{q_j}S-a_j)+V\nonumber\\
-\frac{\lambda^2}{2}\Big(g^{ij}\frac{\partial_{q_i}\partial_{q_j}R}{R}+\partial_{q_i}g^{ij}\frac{\partial_{q_j}R}{R}\Big)=0.
\label{HJM particle in potentials}
\end{eqnarray} 

We have thus a pair of coupled equations (\ref{FPE particle in potentials}) and (\ref{HJM particle in potentials}) which are parameterized by $\lambda$. Recall that this pair of equations are valid in a microscopic time interval of length $\tau_{\xi}$ during which the magnitude of the envelope of $\xi$ is constant while the sign of $\xi$ changes randomly with equal probability. Moreover, recall also that the sign of $\lambda$ is always the same as the sign of $\xi$. Noting this, averaging Eq. (\ref{FPE particle in potentials}) for the cases $\pm\xi$, thus is also over $\pm\lambda$, one has, by virtue of Eqs. (\ref{God's fairness}) and (\ref{quantum phase symmetry}), 
\begin{equation}
\partial_t\Omega+\partial_{q_i}\Big(\widetilde{g^{ij}}(\partial_{q_j}S-a_j)\Omega\Big)+\frac{|\lambda|}{2}\partial_{q_i}(\Delta g^{ij}\partial_{q_j}\Omega)=0, 
\label{QCE particle in potentials}
\end{equation}
where $\widetilde{g_{ij}}$ and $\Delta g_{ij}$ are defined as
\begin{equation}
\widetilde{g^{ij}}\doteq\frac{g^{ij}(\xi)+g^{ij}(-\xi)}{2},\hspace{2mm}\Delta g^{ij}\doteq\frac{g^{ij}(\xi)-g^{ij}(-\xi)}{2}.
\label{fluctuation of metric 0} 
\end{equation} 
Similarly, averaging Eq. (\ref{HJM particle in potentials}) over the cases $\pm\xi$ one gets 
\begin{eqnarray}
\partial_tS+\frac{\widetilde{g^{ij}}}{2}(\partial_{q_i}S-a_i)(\partial_{q_j}S-a_j)+V\nonumber\\
-\frac{\lambda^2}{2}\Big(\widetilde{g^{ij}}\frac{\partial_{q_i}\partial_{q_j}R}{R}+\partial_{q_i}\widetilde{g^{ij}}\frac{\partial_{q_j}R}{R}\Big)=0.
\label{HJM particle in potentials metric}
\end{eqnarray}
We have thus a pair of coupled equations (\ref{QCE particle in potentials}) and (\ref{HJM particle in potentials metric}) which are now parameterized by a constant $|\lambda|$, valid during a microscopic time interval of length $\tau_{\xi}$ characterized by a constant $\|\xi\|$. 

Next, let us assume that the metric is fluctuating weakly around a classical background $g_C^{ij}(q)$ as follows 
\begin{eqnarray}
g^{ij}(q;\xi)=g_C^{ij}(q)+f^{ij}(\xi),\hspace{20mm}\nonumber\\
\mbox{with}\hspace{2mm}f^{ij}(\xi)=-f^{ij}(-\xi)\hspace{2mm}\mbox{and}\hspace{2mm}|f^{ij}(\xi)|\sim o(|\xi|). 
\end{eqnarray}
Hence, $f^{ij}(\xi)$ is a global deviation from the classical metric $g_C^{ij}(q)$ with anti-symmetric property. In this case, for sufficiently small $\|\xi\|$ one has 
\begin{equation}
\widetilde{g^{ij}}=g_C^{ij},\hspace{2mm}\mbox{and}\hspace{2mm}\Delta g^{ij}\approx 0.
\label{fluctuation of metric} 
\end{equation}
so that the pair of Eqs. (\ref{QCE particle in potentials}) and (\ref{HJM particle in potentials metric}) reduce to
\begin{eqnarray}
\partial_t\Omega+\partial_{q_i}\Big(g_C^{ij}(\partial_{q_j}S-a_j)\Omega\Big)=0,\hspace{6mm}\nonumber\\ 
\partial_tS+\frac{g_C^{ij}}{2}(\partial_{q_i}S-a_i)(\partial_{q_j}S-a_j)+V\nonumber\\
-\frac{\lambda^2}{2}\Big(g_C^{ij}\frac{\partial_{q_i}\partial_{q_j}R}{R}+\partial_{q_i}g_C^{ij}\frac{\partial_{q_j}R}{R}\Big)=0.
\label{Modified Madelung equation particle in potentials}
\end{eqnarray}
 
Next, since $|\lambda|$ is non-vanishing, one can define the following complex-valued function:
\begin{equation}
\Psi\doteq \sqrt{\Omega}\exp\Big(i\frac{S}{|\lambda|}\Big). 
\label{general wave function}
\end{equation}
Using $\Psi$, recalling the assumption that $|\lambda|$ is constant during the time interval of interest, the pair of equations in (\ref{Modified Madelung equation particle in potentials}) can then be recast into the following equation: 
\begin{equation}
i|\lambda|\partial_t\Psi=\hat{H}^q_{|\lambda|}\Psi,
\label{generalized Schroedinger equation particle in potentials} 
\end{equation} 
where $\hat{H}^q_{|\lambda|}$ is defined as 
\begin{equation}
\hat{H}^q_{|\lambda|}\doteq \frac{1}{2}(\hat{p}^q_{|\lambda|_i}-a_i)g_C^{ij}(q)(\hat{p}^q_{|\lambda|_j}-a_j)+V,
\label{lambda-parameterized QH}  
\end{equation}  
with $\hat{p}^q_{|\lambda|}\doteq-i|\lambda|\partial_q$.  
 
Let us then consider a specific case when $|\lambda|$ is given by the reduced Planck constant $\hbar$, so that the exponential distribution of the deviation from infinitesimal stationary action $P_S(dS|dA)$ of Eq. (\ref{exponential distribution of DISA}) is stationary in time with average
\begin{equation}
\hbar/2. 
\label{the P in Q}
\end{equation}
Let us further assume that the fluctuation of $\xi$ around its vanishing average is sufficiently narrow. One may therefore approximate $\Omega(q,\xi;t)$ and $S(q;t,\xi)$ by the corresponding zeroth order terms of their Taylor expansion, respectively denoted by $\rho_Q(q;t)$ and $S_Q(q;t)$. In this specific case, the zeroth order approximation of Eq. (\ref{generalized Schroedinger equation particle in potentials}) gives the Schr\"odinger equation 
\begin{eqnarray}
i\hbar\partial_t\Psi_Q(q;t)=\hat{H}^q_{\hbar}\Psi_Q(q;t),\hspace{5mm}\nonumber\\
\Psi_Q(q;t)\doteq\sqrt{\rho_Q(q;t)}e^{\frac{i}{\hbar}S_Q(q;t)},\hspace{0mm}
\label{Schroedinger equation particle in potentials} 
\end{eqnarray} 
where $\hat{H}^q_{\hbar}$ is just the quantum Hamiltonian in position representation
\begin{equation}
\hat{H}^q_{\hbar}=\frac{1}{2}(\hat{p}^q_i-a_i)g_C^{ij}(q)(\hat{p}^q_j-a_j)+V, 
\label{quantum Hamiltonian particle in potentials}
\end{equation}
with $\hat{p}^q\doteq\hat{p}^q_{\hbar}=-i\hbar\partial_q$ is the quantum mechanical Hermitian momentum operator. From Eq. (\ref{Schroedinger equation particle in potentials}), one can see that the Born's statistical interpretation of wave function is valid by construction
\begin{equation}
\rho_Q(q;t)=|\Psi_Q(q;t)|^2.
\label{Born's statistical interpretation}  
\end{equation} 

Further, recall again that fixing $\|\xi\|$, Eq. (\ref{classical velocity field HPF particle in potentials}) is valid only within a time interval of length $\tau_{\xi}$ during which the sign of $\xi$ is fluctuating randomly with equal probability. It is therefore natural to define an effective velocity as $\widetilde{\dot{q}}(|\xi|)\doteq(\dot{q}(\xi)+\dot{q}(-\xi))/2$. Inserting Eq. (\ref{classical velocity field HPF particle in potentials}), recalling the fact that the sign of $\xi$ is the same as that of $\lambda$, and noting Eqs. (\ref{God's fairness}), (\ref{quantum phase symmetry}) and (\ref{fluctuation of metric}), one has, in the lowest order approximation 
\begin{equation}
\widetilde{\dot{q}}^i\approx\widetilde{g^{ij}}(\partial_{q_j}S_Q-a_j)=g_C^{ij}(\partial_{q_j}S_Q-a_j).  
\label{Bohmian velocity}
\end{equation}  

One can also show that the average of relevant physical quantities $O(q,p)$ at most quadratic in momentum over the distribution of the configuration $\Omega(q,\xi)$ is given by the quantum mechanical average of the corresponding Hermitian operators over a wave function \cite{AgungSMQ6,AgungSMQ8}. To see this, without losing generality, let us calculate the average energy of a particle of mass $m$ subjected to a scalar potential $V(q)$. The metric is thus given by $g^{ij}=\delta_{ij}/m$, where $\delta_{ij}$ is the Kronecker delta, assumed to be weakly fluctuating chaotically around its classical background $\delta_{ij}/m_C$ satisfying Eq. (\ref{fluctuation of metric}). One obtains 
\begin{eqnarray}
\langle H\rangle\doteq\int dqd\xi H(q,p)\Omega(q,\xi)=\int dq d\xi\Big(\frac{p^2}{2m}+V\Big)\Omega\hspace{15mm}\nonumber\\
=\int dqd\xi\Big[\frac{1}{2m}\Big((\partial_qS)^2+\lambda\partial_qS\frac{\partial_q\Omega}{\Omega}+\Big(\frac{\lambda}{2}\frac{\partial_q\Omega}{\Omega}\Big)^2\Big)+V\Big]\Omega\nonumber\\
\approx \int dq\Big(\frac{(\partial_qS_Q)^2}{2m_C}-\frac{\hbar^2}{2m_C}\frac{\partial_q^2\sqrt{\rho_Q}}{\sqrt{\rho_Q}}+V\Big)\rho_Q\nonumber\\
=\int dq\Psi_Q^*\hat{H}_{\hbar}^q\Psi_Q\doteq\langle\Psi_Q|\hat{H}^q_{\hbar}|\Psi_Q\rangle.
\label{statistical average=quantum average} 
\end{eqnarray} 
Here in the second equality we have inserted the upper equation of (\ref{fundamental equation rederived}), in the third approximate equality we have used Eqs. (\ref{fluctuation decomposition}) and (\ref{fluctuation of metric}), noting the fact that by construction the sign of $\lambda$ is always the same as that of $\xi$ and Eqs. (\ref{God's fairness}) and (\ref{quantum phase symmetry}) to eliminate the second term, counted only the zeroth order terms and considered the case when $|\lambda|=\hbar$, and in the last equality we have used the definition of the wave function given in Eq. (\ref{Schroedinger equation particle in potentials}).  

Let us remark that all the statistical results above are obtained by averaging over the fluctuation of $\xi$ in a time scale $dt$. In other words, the information of the system within a time interval $dt$ is traced over. Hence, all the statistical results of the model are not valid for a time scale shorter than $dt$. Since quantum mechanics is recovered as a specific case of the model, the above observation suggests that there is a fundamental microscopic time scale below which quantum mechanics might be no longer applicable. 

Let us mention that there has been many efforts to reconstruct quantum mechanics by deriving the Schr\"odinger equation from statistical models \cite{Nelson,Garbaczewski,tHooft,Markopoulou,dela Pena 1}, assuming that quantum fluctuation is physically and objectively real. One of the important feature of the present derivation is that it is derived from a statistical deterministic model that is {\it not only} local satisfying the principle of Locality, but, as argued in Refs. \cite{AgungSMQ6,AgungSMQ8,AgungSMQ9}, is {\it singled out uniquely} by the later. 
 
\subsection{Measurement of angular momentum, Born's rule and no-signaling}
 
We have also applied the statistical model to the measurement of angular momentum, reproducing the prediction of quantum mechanics \cite{AgungSMQ7}. See also Ref. \cite{AgungSMQ9} for the application of the model to Stern-Gerlach experiment. To do this, we have considered two particles interacting via the von-Neumann classical Hamiltonian
\begin{equation}
H_I=g(\xi)l_{z_1}{p}_2,\hspace{2mm}\mbox{with}\hspace{2mm}l_{z_1}=x_1{p}_{y_1}-y_1{p}_{x_1},   
\label{classical Hamiltonian measurement angular momentum}
\end{equation}  
where $g(\xi)$ is a coupling constant assumed to depend on $\xi$, fluctuating weakly around its classical value $g_C$ admitting the following decomposition
\begin{eqnarray}
g(\xi)=g_C+f(\xi),\hspace{30mm}\nonumber\\
\mbox{with}\hspace{2mm}f(\xi)=-f(-\xi)\hspace{2mm}\mbox{and}\hspace{2mm}|f(\xi)|\sim o(|\xi|),
\label{coupling fluctuation}
\end{eqnarray} 
and $l_{z_1}$ is the $z-$angular momentum of the first particle. Classically, the above  interaction-Hamiltonian with $g$ replaced by $g_C$ can be used to model the measurement of the angular momentum of the first particle by regarding the position of the second particle as the pointer reading. Namely, the final position of the second particle is determined by the value of the angular momentum of the first particle {\it prior} to interaction, so that the latter can be {\it inferred} by looking at the former.  

Applying the statistical model, one first will obtain the following Schr\"odinger equation, as the lowest order approximation \cite{AgungSMQ7}:
\begin{eqnarray}
i\hbar\partial_t\Psi_Q(q;t)=\hat{H}^q_I\Psi_Q(q;t),\hspace{5mm}\nonumber\\
\Psi_Q(q;t)\doteq\sqrt{\rho_Q(q;t)}e^{\frac{i}{\hbar}S_Q(q;t)}.\hspace{0mm} 
\label{Schroedinger equation measurement of angular momentum}
\end{eqnarray}  
Here $\hat{H}^q_I$ is the corresponding quantum Hamiltonian, a differential operator defined as 
\begin{equation}
{\hat H}^q_I\doteq g_C{\hat l}^q_{z_1}{\hat p}^q_2,
\label{Hamiltonian operator angular momentum}
\end{equation} 
where $\hat{p}^q_i\doteq-i\hbar\partial_{q_i}$, $i=1,2$ is the quantum mechanical linear momentum operator referring to the $i-$particle and ${\hat l}^q_{z_1}\doteq x_1{\hat p}_{y_1}-y_1{\hat p}_{x_1}$ is the $z-$part of the quantum mechanical angular momentum operator of the first particle.

One can then show that starting from a separable wave function $\Psi_Q(q;0)=\psi_0(q_1)\varphi_0(q_2)$, and expanding the initial wave function of the first particle as $\psi_0(q_1)=\sum_lc_l\phi_l$, where $\{\phi_l\}$, $l=1,2,\dots$ is the complete set of orthonormal eigenfunctions of ${\hat l}^q_{z_1}$, one has, at time $t$, 
\begin{equation}
\Psi_Q(q_1,q_2;t)=\sum_lc_l\phi_l(q_1)\varphi_0(q_2-g_C\omega_lt), 
\label{entangled system-apparatus}
\end{equation}
where $\omega_l$ is the eigenvalue of ${\hat l}^q_{z_1}$ with the associated eigenfunction $\phi_l$. One can see that if $\varphi_0(q_2)$ is sufficiently narrow, then for sufficiently large $g_C$ and $t$, $\{\varphi_l(q_2;t)\doteq\varphi_0(q_2-g_C\omega_lt)\}$ are not overlapping for different values of $l$. 

One then proceeds as follows \cite{AgungSMQ7}. First, to have a physically and operationally smooth quantum-classical correspondence, one must let $q_2(t)$ has the same physical and operational status as the underlying classical mechanical system: namely, it must be regarded as the pointer of the measurement, the reading of our experiment, or the `hidden variable' that determines the `outcome' of measurement. One may then {\it infer} that the outcome of a single measurement event corresponds to the packet $\varphi_l(q_2;t)$ whose support is actually entered by the apparatus particle. Namely, if $q_2(t)$ belongs to the spatially localized support of $\varphi_l(q_2;t)$, then we {\it operationally} admit that the result of measurement is given by $\omega_l$, the eigenvalue of $\hat{l}^q_{z_1}$ whose corresponding eigenfunction $\phi_l(q_1)$ is correlated with $\varphi_l(q_2;t)$. The probability that the measurement yields $\omega_l$ is thus equal to the relative frequency that $q_2(t)$ enters the support of $\varphi_l(q_2;t)$ in a large (in principle infinite) number of identical experiments, which can be easily shown to be given by $|c_l|^2$  \cite{AgungSMQ7,
AgungSMQ9}, the Born's rule.   

We have thus argued within the statistical model that measurement of angular momentum can be described as a specific type of physical interaction. Namely, the cases when there is no measurement and when there is a measurement are treated in a unified way within the statistical model, satisfying the same dynamical and statistical law given by the exponential distribution of deviation from infinitesimal stationary action of Eq. (\ref{exponential distribution of DISA}), necessitating no external concept. In particular, there is no need for instantaneous wave function collapse. The wave function of the whole system + apparatus still follows the unitary Schr\"odinger equation of (\ref{Schroedinger equation measurement of angular momentum}) and the local `click' in the detection (measurement) event is provided by the configuration of the apparatus $q_2(t)$. 

The above mechanism of measurement is similar to the no-collapse pilot-wave theory \cite{pilot-wave theory}, in which the configuration of the system is also regarded as the hidden variables which together with the wave function describe completely the state of the system. However, unlike pilot-wave theory which is based on the fundamental assumption that the wave function is a {\it real-physical field} (psi-ontic), the wave function in the statistical model is an artificial mathematical construct with no fundamental physical ontology (psi-epistemic). It is well-known that a physical wave function living in {\it configuration space} rather than in ordinary space leads to rigid nonlocality. By contrast, the present statistical model satisfies the separability condition of Eq. (\ref{principle of Local Causality}) so that it is {\it objectively local}. In this sense, Eq. (\ref{Bohmian velocity}) can {\it not} be regarded as a causal-dynamical guidance relation as in pilot-wave theory, rather it is a kinematical relation. Moreover, unlike pilot-wave theory in which the quantum dynamics and kinematics are postulated, and so is the additional guidance relation, in the statistical model, they are derived from a set of general principles.    

Now let us consider a Bell-type experiment where one is interested in the statistical correlation between a set of pairs of spacelike separated measurement events.  Since the statistics of {\it any} events, whether they are measurement events or not, must satisfy the objective separability condition of Eq. (\ref{principle of Local Causality}) when they are separated by spacelike interval describing the independence of one event from the other in the pair, then the statistics of the measurement outcomes at one wing of the Bell-type experiment must be independent of the local parameter of the other wing \cite{AgungSMQ9}. This prohibits an experimenter at one of the wing to send signal, by varying the local parameters at his/her hand, to the experimenter at the other wing, thus the no-signaling.     

Let us mention that standard quantum mechanics has also been shown to respect the principle of no-signaling \cite{Eberhard no-signaling,GRW no-signaling,Jarret no-signaling,Shimony 0}. Moreover, a nonlinear modification of its dynamical law \cite{Weinberg nonlinearity} may lead to signaling \cite{Gisin signaling,Polchinski signaling}, contradicting the theory of relativity. We shall show in the next sections that the above no-signaling statistical model leads to a modification of quantum dynamics while preserving its celebrated linearity property.    

\section{A time evolution with chaotic parameter} 
 
Hence, we have a space (landscape) of possible theories, all follow from a chaotic fluctuation of infinitesimal stationary action with a distribution that is given by the exponential law Eq. (\ref{exponential distribution of DISA}), parameterized by $|\lambda(t)|$ which in general is fluctuating chaotically in a microscopic time scale  $\tau_{\lambda}$. The Schr\"odinger equation of (\ref{Schroedinger equation particle in potentials}) with the Born's statistical interpretation of wave function of Eq. (\ref{Born's statistical interpretation}) are then valid approximately for a specific theory in the theoryspace corresponding to a specific value of parameter $|\lambda|=\hbar$ for all the time, so that the average of the deviation from infinitesimal stationary action distributed according to the exponential law of Eq. (\ref{exponential distribution of DISA}) is given by $\hbar/2$. This is the case when $\tau_{\lambda}$ is {\it infinite}. If quantum mechanics is exact, then it is of great interest to have a set of physical axioms which uniquely selects $|\lambda|=\hbar$. In particular, the axioms must also explain the physical origin of the numerical value of $\hbar$ as observed in experiment. Or, one may conceive a deeper level theory which leads effectively to the present statistical model so that the value of $\hbar$ is computable. 

With the absence of theoretical justification for $|\lambda|=\hbar$ in the statistical model, it is then not unreasonable to  assume that quantum mechanics is {\it not} exact. Below we shall assume that there are physical situations in which $|\lambda(t)|$ might be fluctuating randomly around $\hbar$ with a small finite width and a {\it finite} time scale $\tau_{\lambda}$. The practical reason for the speculation that this may be the case is that all experimental results are inevitably limited by the finite accuracy of the measurement devices and the finite spatiotemporal accessibility of the physical phenomena under study, and is also bounded by the scale of energy involved, etc. Moreover, theoretically, within the statistical model discussed above, it seems physically very unlikely that Nature is discontinuous at $|\lambda|=\hbar$ so that quantum mechanics becomes an `island' in the theory space \cite{Aaronson's island} corresponding to the case when $|\lambda|=\hbar$ in the statistical model, and the other cases when $|\lambda|\neq\hbar$ are unrealizable for some unknown reasons. In other words, there is a priori no compelling physical (non-anthropic) reason, except to fit the experimental results under some allowable measurement uncertainty, why the distribution of deviation from the infinitesimal stationary action given by the exponential law of Eq. (\ref{exponential distribution of DISA}) is stationary for all time with an average that is equal exactly to $\hbar/2$ in any physical phenomena. In the absence of sufficient reason, it is then advisable to assume more general cases that $|\lambda|$ can have values different from $\hbar$.    
 
We shall proceed to consider the implication of the possibility that $|\lambda|$ is fluctuating effectively randomly around $\hbar$ in a finite microscopic time scale $\tau_{\lambda}$. Hence, we shall regard the $|\lambda|-$parameterized Schr\"odinger equation of (\ref{generalized Schroedinger equation particle in potentials}) as a straightforward generalization of the usual Schr\"odinger equation of (\ref{Schroedinger equation particle in potentials}). Since the latter is a specific case of the former when $|\lambda|$ is constant for all the time given by the reduced Planck constant $\hbar$, then Eq. (\ref{generalized Schroedinger equation particle in potentials}) may be physically interpreted to generalize quantum dynamics by allowing the Planck `constant' fluctuates with time. In other words, the proposed modification is equivalent to the assumption that the currently suggested value of the reduced Planck constant is possibly a very accurate extrapolation of the random (chaotic) parameter $|\lambda(t)|$ of the statistical model. 
  
To check the above assumption directly, one needs to make a detailed statistical analysis on the uncertainty of the measurement of Planck constant, the latest value of which is given by $\propto 10^{-8}$ \cite{Steiner PlanckC}. Instead of doing this, we shall show in the next section that such a random fluctuation leads effectively to a non-unitary time evolution implying a universal intrinsic mechanism of decoherence in energy space in the macroscopic regime. A similar remark on possible random fluctuation of Planck constant is suggested by Calogero \cite{Calogero} in his attempt to provide a cosmological origin of the numerical value of Planck constant within the framework of a general class of stochastic models of quantization. Calogero also hinted with no detail elaboration that such a stochastic correction may lead to an ultra-weak violation of time-reversal symmetry, thus a violation of the unitary time evolution of quantum mechanics. Let us emphasize before proceeding that the modification of quantum dynamics proposed in the present paper is {\it naturally} suggested by the physical interpretation of the Planck constant that appears in the Schr\"odinger equation within the statistical model as the average of deviation from classical mechanics in a microscopic time scale.     

Let us again assume that $\Omega(q,\xi;t)$ and $S(q;t,\xi)$ can be approximated by the zeroth order terms of the corresponding Taylor expansions around the vanishing average of $\xi$, denoted by  $\rho_Q(q;t)$ and $S_Q(q;t)$, respectively. Let us then consider the case when $t$ may be much larger than $\tau_{\lambda}$. During the time interval of interest, $|\lambda(t)|$ may thus fluctuate chaotically following a probability distribution $P(\lambda(t))=P(-\lambda(t))$, and $P_S(dS|dA)$ of Eq. (\ref{exponential distribution of DISA}) is thereby {\it not} stationary. In this case, from Eq. (\ref{fundamental equation 0}), the probability density that the system has a configuration $q$ at time $t$ must be conditioned on the functional values of $|\lambda(t)|$ up to time $t$, thus should be denoted as $\rho_Q(q;t|[\lambda(t)])$. $([\lambda(t)])$ is here used to denote a functional dependence on $\lambda(t)$. The zeroth order term of the wave function at time $t$ thus takes the following form:  
\begin{equation}
\Psi_Q(q;t\big|[\lambda(t)])\doteq\sqrt{\rho_Q(q;t\big|[\lambda(t)])}e^{iS_Q/|\lambda(t)|}, 
\label{general wave function 1}
\end{equation}
depending on the values of $|\lambda(t)|$ up to time $t$, and must satisfy the following equation: 
\begin{equation}
i|\lambda(t)|\partial_t\Psi_Q=\hat{H}^q_{|\lambda(t)|}\Psi_Q,
\label{generalized Schroedinger equation particle in potentials 1} 
\end{equation} 
which should again be regarded as a generalization of the Schr\"odinger equation of (\ref{Schroedinger equation particle in potentials}), valid even when $t$ is much larger than $\tau_{\lambda}$.  

Let us note before proceeding that fixing $|\lambda|$ for a time interval less than $\tau_{\lambda}$, instead of Eq. (\ref{statistical average=quantum average}), the average energy of a particle of mass $m$ is given by  $\langle H\rangle=\langle\Psi_Q|\hat{H}^q_{|\lambda|}|\Psi_Q\rangle$: Eq. (\ref{statistical average=quantum average}) is a specific case when $|\lambda|=\hbar$. This can be generalized to any physical quantities $O(q,p)$ up to second order in momentum to have $\langle O\rangle=\langle\Psi_Q|\hat{O}^q_{|\lambda|}|\Psi_Q\rangle$, where $\hat{O}^q_{|\lambda|}$ is the corresponding Hermitian operator whose form for $|\lambda|=\hbar$ is just the quantum mechanical Hermitian operator corresponding the physical quantity $O$. Noting this fact, taking the derivative with respect to time, and using the generalized Schr\"odinger equation of Eq. (\ref{generalized Schroedinger equation particle in potentials 1}), one obtains the Ehrenfest theorem
\begin{eqnarray}
\frac{d}{dt}\langle O\rangle=\frac{1}{i|\lambda|}\langle\Psi_Q|[\hat{O}^q_{|\lambda|},\hat{H}^q_{|\lambda|}]|\Psi_Q\rangle+\langle\Psi_Q|\partial_t\hat{O}^q_{|\lambda|}|\Psi_Q\rangle, 
\end{eqnarray}   
where for any two operators $\hat{O}_1$ and $\hat{O}_2$,  $[\hat{O}_1,\hat{O}_2]\doteq\hat{O}_1\hat{O}_2-\hat{O}_2\hat{O}_1$. In particular, for $O=p$, as shown in the previous section one has $\hat{O}^q_{|\lambda|}=-i|\lambda|\partial_q$ so that the above equation reduces into  
\begin{equation}
\frac{d\langle p\rangle}{dt}=\langle\Psi_Q|(-\partial_qV)|\Psi_Q\rangle=\int dq(-\partial_qV)\rho_Q=\langle(-\partial_qV)\rangle. 
\end{equation}
Notice that the above equation is no longer parameterized by $|\lambda|$. 

Below, since only the absolute value of $\lambda$ that matters, for notational simplicity, we shall sometime write $\lambda$ in place of $|\lambda|$ with the understanding that its negative counterpart may occur with equal probability. Applying the conventional probability theory, the probability density that the configuration of the system is $q$ at time $t$ is therefore related to the wave function defined in Eq. (\ref{general wave function 1}) as  
\begin{eqnarray}
\rho(q;t)=\int D[\lambda(t)]\rho_Q(q;t\big|[\lambda(t)])P([\lambda(t)])\hspace{00mm}\nonumber\\
=\int D[\lambda(t)]\big|\Psi_Q(q;t\big|[\lambda(t)])\big|^2P([\lambda(t)]),
\label{generalized Born's rule}
\end{eqnarray}
where the functional integration $\int D[\lambda(t)]\dots$ is over all possible functional forms of $\lambda(t)$ up to time $t$ of interest. One can see that in the case when $\lambda(t)=\pm\hbar$ for all the time with equal probability then Eq. (\ref{generalized Born's rule}) reduces to the Born's prescription of Eq. (\ref{Born's statistical interpretation}). In this sense, the former should thus be regarded as a natural generalization of the latter. 

Now, for later purpose, let us develop an abstract Hilbert space formalism as in standard quantum mechanics. First, let us assume that $\{|q\rangle\}$ spans the Hilbert space and associate to each wave function $\Psi_Q(q;t|[\lambda(t)])$, parameterized by the value of $\lambda(t)$ up to time $t$, a (state) vector in the Hilbert space as $|\Psi_Q(t|[\lambda(t)])\rangle\doteq\int dq\Psi_Q(q;t|[\lambda(t)])|q\rangle$. The dual basis is denoted as $\langle q|$ and a dual vector is defined as $\langle\Psi_Q(t|[\lambda(t)])|\doteq\int dq\Psi_Q^*(q;t|[\lambda(t)])\langle q|$. The basis is orthonormal satisfying $\langle q|q'\rangle=\delta(q-q')$, where $\langle\diamond|\circ\rangle$ is the inner product between $|\diamond\rangle$ and $|\circ\rangle$. The inner product between two vectors, parameterized by the same form of $\lambda(t)$ up to time $t$, is therefore given by $\langle\Psi_Q(t|[\lambda(t)])|\Phi_Q(t|[\lambda(t)])\rangle=\int dq\Psi_Q^*(q;t|[\lambda(t)])\Phi_Q(q;t|[\lambda(t)])$. 

Let us also introduce a representation free quantum Hamiltonian $\hat{H}_{\lambda(t)}$ whose operation on $|\Psi_Q(t|[\lambda(t)])\rangle$ is defined as $\hat{H}_{\lambda(t)}|\Psi_Q(t|[\lambda(t)])\rangle\doteq\int dq \hat{H}_{\lambda(t)}^q\Psi_Q(q;t|[\lambda(t)])|q\rangle$. Hence, if $\Psi_Q(q;t|[\lambda(t)])$ satisfies the generalized Schr\"odinger equation of (\ref{generalized Schroedinger equation particle in potentials 1}), then $|\Psi_Q(t|[\lambda(t)])\rangle$ satisfies the following equation:
\begin{equation}
i\lambda(t)\frac{d}{dt}|\Psi_Q(t|[\lambda(t)])\rangle=\hat{H}_{\lambda(t)}|\Psi_Q(t|[\lambda(t)])\rangle. 
\label{generalized Schroedinger equation free representation}
\end{equation}  
Formally, assuming that $\hat{H}_{\lambda}$ does not explicitly depend on time, it can be solved to give 
\begin{equation}
|\Psi_Q(t|[\lambda(t)])\rangle=e^{-i\int^t_{t_0}dt'\hat{H}_{\lambda(t')}/\lambda(t')}|\Psi_Q(t_0|[\lambda(t_0)])\rangle, 
\label{general solution}
\end{equation}
where $|\Psi_Q(t_0|[\lambda(t_0)])\rangle$ is the initial state which depends on the value of $\lambda$ up to the initial time $t_0$. 

By construction, one also has, for a fixed $\lambda$
\begin{equation}
\hat{H}_{\lambda}|\phi_i(\lambda)\rangle=E_i(\lambda)|\phi_i(\lambda)\rangle,\hspace{2mm}i=0,1,2,\dots,
\label{random spectrum operator form}
\end{equation}
where $|\phi_i(\lambda)\rangle\doteq\int dq\phi_i(q|\lambda)|q\rangle$, and $\{E_i(\lambda)\}$ and $\{\phi_i(q|\lambda)\}$ are the set of eigenvalues and the corresponding eigenfunctions of $\hat{H}^q_{\lambda}$ which can be obtained by replacing $\hbar$ with $\lambda$ to those of the corresponding quantum Hamiltonian $\hat{H}_{\hbar}^q$ in the position representation. For a given value of $\lambda$, the set of the eigenvectors $\{|\phi_i(\lambda)\rangle\}$ are orthonormal $\langle\phi_i(\lambda)|\phi_j(\lambda)\rangle=\delta_{ij}$ and complete $\sum_i|\phi_i(\lambda)\rangle\langle\phi_i(\lambda)|=\hat{I}$, where $\hat{I}$ is the identity operator. It thus spans the Hilbert space. Hence the set of energy eigen basis is fluctuating with time. Any state of the system can be expanded as the superposition of the eigenfunctions of $\hat{H}_{\lambda}$ for any value of the parameter $\lambda$, 
\begin{eqnarray} 
|\Psi_Q(t|[\lambda(t)])\rangle=\sum_mc_m|\phi_m(\lambda)\rangle,\nonumber\\
\mbox{with}\hspace{2mm}c_m=\langle\phi_m(\lambda)|\Psi_Q(t|[\lambda(t)])\rangle.
\end{eqnarray}   

Using the above formalism, the generalized Born's prescription of Eq. (\ref{generalized Born's rule}) can thus be written as 
\begin{eqnarray}
\rho(q;t)=\int D[\lambda(t)]\langle q|\Psi_Q(t\big|[\lambda(t)])\rangle\langle\Psi_Q(t\big|[\lambda(t)])|q\rangle\nonumber\\
\times P([\lambda(t)]).\hspace{20mm}
\label{generalized Born's rule Hilbert}
\end{eqnarray}
Let us define a $\lambda-$parameterized density matrix as in standard quantum mechanics as 
\begin{equation}
\hat{\rho}(t\big|[\lambda(t)])=|\Psi_Q(t\big|[\lambda(t)])\rangle\langle\Psi_Q(t\big|[\lambda(t)])|. 
\label{lambda-parameterized density matrix}
\end{equation}
Using the density matrix, the generalized Born's prescription of Eq. (\ref{generalized Born's rule Hilbert}) can thus be written as 
\begin{eqnarray}
\rho(q;t)=\langle q|\widetilde{\rho}(t)|q\rangle,
\label{generalized Born's rule density matrix}
\end{eqnarray}
where $\widetilde{\rho}(t)$ is defined as 
\begin{eqnarray}
\widetilde{\rho}(t)\doteq\int D[\lambda(t)] \hat{\rho}(t\big|[\lambda(t)]) P([\lambda(t)]),
\label{lambda-averaged density matrix}
\end{eqnarray}
namely, it is the average of the density matrix over all possible functional form of $\lambda(t)$ up to time $t$. 

To further evaluate the above functional integration, let us make the following approximation. Let us discretize the total macroscopic time of interest $t$ into $n$ finite intervals of length $\tau_{\lambda}$ and attribute to each interval a value of $|\lambda|$ randomly drawn from a probability distribution. We have thus a dynamical process $\{|\lambda_1|,|\lambda_2|,\dots,|\lambda_n|\}$, where $|\lambda|=|\lambda_k|\neq 0$ is held constant for each interval of time $t\in[(k-1)\tau_{\lambda},k\tau_{\lambda})$, $k=1,2\dots,n$. Let us further assume that $\lambda(t)$ is sufficiently chaotic so that $\{|\lambda_k|\}$ are independent of each other and follow identical distribution $P(|\lambda|)$. Hence, the value of $|\lambda|$ at any time interval is assumed to be independent from its values in the past, which is only approximately valid when $|\lambda(t)|$ is sufficiently chaotic. As shown in the previous section, the case when $\lambda_k=\pm\hbar$ with equal probability for all $k=1,2,\dots,n$ so that $P_S(dS|dA)$ of Eq. (\ref{exponential distribution of DISA}) is stationary, leads to a unitary time evolution given by the Schr\"odinger equation. Below we shall go beyond this specific case by allowing $|\lambda_k|$ fluctuates randomly around $\hbar$. 

Let us first consider an interval of time during which $\lambda$ is constant. Then, if $\hat{H}_{\lambda}$ does not depend explicitly on time, from Eq. (\ref{general solution}) one has 
\begin{eqnarray}
|\Psi_Q(t|[\lambda(t)])\rangle=\hat{U}_{\lambda}(t-t_0)|\Psi_Q(t_0|[\lambda(t_0)])\rangle,\nonumber\\
\mbox{where}\hspace{2mm}\hat{U}_{\lambda}(t-t_0)\doteq\exp(-i\hat{H}_{\lambda}(t-t_0)/\lambda). 
\label{unitary stochastic time evolution operator}
\end{eqnarray}
Now let us consider the dynamical evolution within a time interval $t\in [(k-1)\tau_{\lambda},k\tau_{\lambda})$ so that $\lambda=\lambda_k$. From Eq. (\ref{unitary stochastic time evolution operator}), one gets
\begin{eqnarray}
|\Psi_Q(t|[\lambda(t)])\rangle=\hat{U}_{\lambda_k}(t-(k-1)\tau_{\lambda})\hspace{30mm}\nonumber\\
\circ|\Psi_Q((k-1)\tau_{\lambda}|[\lambda((k-1)\tau_{\lambda})])\rangle,\hspace{2mm}k=1,2,\dots,n.  
\label{one time step}  
\end{eqnarray}
Applying the above equation consecutively for all intervals of time, one thus obtains 
\begin{eqnarray}
|\Psi(t|[\lambda(t)])\rangle=\hat{U}_{\lambda_k}(t-(k-1)\tau_{\lambda})\circ\dots\nonumber\\
\dots\circ\hat{U}_{\lambda_2}(\tau_{\lambda})\circ\hat{U}_{\lambda_1}(\tau_{\lambda})|\Psi(t_0|[\lambda(t_0)])\rangle.\hspace{0mm}  
\label{n time step} 
\end{eqnarray}

Again, from Eq. (\ref{unitary stochastic time evolution operator}), if $\lambda=\hbar$, the time evolution operator for different intervals $t\in[(k-1)\tau_{\lambda},k\tau_{\lambda})$, $k=1,\dots,n$, are now identical given by $\hat{U}_{\hbar}(\Delta t)$. The evolution of the system for the whole period of time of interest $t=n\tau_{\lambda}$ is thus governed by the standard quantum mechanical time evolution operator $
\hat{U}_{\hbar}(t)=\prod_{i=1}^n\hat{U}_{\hbar}(\tau_{\lambda})=\prod_{i=1}^n e^{-\frac{i}{\hbar}\hat{H}_{\hbar}\tau_{\lambda}}=e^{-\frac{i}{\hbar}\hat{H}_{\hbar}t}$. 
Hence, as expected, within the above approximation, the quantum mechanical unitary time evolution operator is a special case when $|\lambda|=\hbar$ for all intervals of time.  
 
Let us investigate the case when $\lambda$ is fluctuating randomly around the vicinity of $\hbar$ with a finite yet small width so that $P_S(dS|dA)$ of Eq. (\ref{exponential distribution of DISA}) is not stationary. First, using Eq. (\ref{n time step}), the $\lambda-$parameterized density matrix of Eq. (\ref{lambda-parameterized density matrix}) reads  
\begin{eqnarray}
\hat{\rho}(t|[\lambda(t)])=\hat{U}_{\lambda_n}(\tau_{\lambda})\circ\dots\circ\hat{U}_{\lambda_2}(\tau_{\lambda})\circ\hat{U}_{\lambda_1}(\tau_{\lambda})\nonumber\\
\circ\hat{\rho}(t_0|[\lambda(t_0)])\circ\hat{U}_{\lambda_1}^{\dagger}(\tau_{\lambda})\circ\hat{U}^{\dagger}_{\lambda_2}(\tau_{\lambda})\circ\dots\circ\hat{U}^{\dagger}_{\lambda_n}(\tau_{\lambda}). 
\end{eqnarray}
Since $P([\lambda(t)])=\prod_i P(\lambda_i)$, the $\lambda-$averaged density matrix of Eq. (\ref{lambda-averaged density matrix}) at time $t=n\tau_{\lambda}$ is thus given by the following path integral over all possible realizations of  $\{\lambda_1,\lambda_2,\dots,\lambda_n\}$
\begin{eqnarray} 
\widetilde{\rho}(t)=\int \prod_id\lambda_i\Big(\prod_k P(\lambda_k)\Big)\hspace{30mm}\nonumber\\
\times\hat{U}_{\lambda_n}(\tau_{\lambda})\circ\dots\circ\hat{U}_{\lambda_2}(\tau_{\lambda})\circ\hat{U}_{\lambda_1}(\tau_{\lambda})\circ\hat{\rho}(t_0|[\lambda(t)])\nonumber\\
\circ\hat{U}_{\lambda_1}^{\dagger}(\tau_{\lambda})\circ\hat{U}^{\dagger}_{\lambda_2}(\tau_{\lambda})\circ\dots\circ\hat{U}^{\dagger}_{\lambda_n}(\tau_{\lambda}). 
\label{path integral for density matrix}
\end{eqnarray}   
Note that here we have made use the assumption that $\{\lambda_k\}$ are independent and identically distributed, which requires that $\tau_{\lambda}$ is much smaller than the time observation $t$, and $\lambda(t)$ is sufficiently chaotic. Let us note however that in reality the values of $\lambda$ at different times should not be strictly uncorrelated since $|\lambda(t)|$ is basically a deterministic function. 

\section{Non-unitary non-dissipative evolution and intrinsic decoherence}

Let us proceed to evaluate the path integral of Eq. (\ref{path integral for density matrix}). To do this, we have to know the distribution of $\lambda$. Let us therefore first infer the distribution of $\lambda$ satisfying the following several reasonable constraints. First, since $\lambda$ is non-vanishing and unbiased satisfying $P(\lambda)=P(-\lambda)$, then it is sufficient to know $P(|\lambda|)$ with $\lambda\neq 0$. Further, since the standard quantum mechanical time evolution operator is reproduced when $\lambda=\pm\hbar$ then to have a smooth `quantum limit', the distribution of $\lambda$ must be such that the former is reproduced as certain limiting case. Namely, somehow one must have $P(|\lambda|)\rightarrow\delta(|\lambda|-\hbar)$. The assumption on a smooth quantum limit also suggests that $P(|\lambda|)$ must have a finite average, which must reduce to $\hbar$ in the quantum limit. Let us also assume that $P(|\lambda|)$ has a finite mean deviation. To summarize, we shall assume that $|\lambda|$ is a non-vanishing positive definite random variable with finite average and deviation. 

Given the above constraints, one can then apply the inference method of maximum entropy principle \cite{Jaynes book} to get the distribution of $|\lambda|$. Namely, among those probability distributions of non-vanishing $|\lambda|$ with finite average and deviation, one chooses the one with maximum Shannon entropy, to have the following symmetric log-normal distribution \cite{Park}:   
\begin{equation}
P(|\lambda|)=\frac{1}{\sqrt{2\pi\sigma^2}|\lambda|}e^{-\frac{(\ln|\lambda|-\mu)^2}{2\sigma^2}}, \hspace{2mm}\lambda \neq 0, 
\label{log-normal pdf random variable} 
\end{equation}
where $\sigma$ is a dimensionless parameter. The location of the modes (peaks) are $\lambda_M=\pm\exp(\mu-\sigma^2)$. $x\doteq\ln|\lambda|$ is normally distributed with mean $\mu$ and width $\sigma$ \cite{log-normal pdf}. In the limit of $\sigma\rightarrow 0$, one has
\begin{equation}
\lim_{\sigma\rightarrow 0}P(|\lambda|)=\delta(|\lambda|-e^{\mu}).
\end{equation} 
Hence, to reproduce the quantum mechanical unitary time evolution as an accurate approximation of the present statistical model, one must identify $\mu$ by the reduced Planck constant as $\mu=\ln\hbar$, and $\sigma^2$ has to be sufficiently small.
  
For the statistical model with the distribution of $P(\lambda)$ given by Eq. (\ref{log-normal pdf random variable}), Eq. (\ref{path integral for density matrix}) can be approximately evaluated as follows. First, let us perform the integration over $\lambda_1$
\begin{equation}
\hat{I}_1\doteq\int d\lambda_1P(\lambda_1)\hat{U}_{\lambda_1}(\tau_{\lambda})\circ\hat{\rho}(0)\circ\hat{U}_{\lambda_1}^{\dagger}(\tau_{\lambda}), 
\label{first integration}
\end{equation}
where $\hat{\rho}(0)\doteq\hat{\rho}(t_0|[\lambda(t_0)])$. To do this, one first expands in energy eigenbasis to get
\begin{eqnarray}
\hat{I}_1=\sum_{mn}\int d\lambda_1P(\lambda_1)\hat{U}_{\lambda_1}(\tau_{\lambda})|\phi_m(\lambda_1)\rangle\langle\phi_m(\lambda_1)|\hat{\rho}(0)\hspace{2mm}\nonumber\\
\circ|\phi_n(\lambda_1)\rangle\langle\phi_n(\lambda_1)|\hat{U}_{\lambda_1}^{\dagger}(\tau_{\lambda})\approx\sum_{mn}|\phi_m(\hbar)\rangle{\rho_Q}_{mn}(0)\langle\phi_n(\hbar)|\nonumber\\
\times\int d\lambda P(\lambda)e^{-\frac{i}{\lambda}(E_m(\lambda)-E_n(\lambda))\tau_{\lambda}}. 
\label{first integration 1}
\end{eqnarray} 
Here, in the second line, we have made an assumption that the fluctuation of $|\phi_n(\lambda)\rangle$ and $\langle\phi_m(\lambda)|\hat{\rho}(0)|\phi_n(\lambda)\rangle$ with respect to $\lambda$ are very smooth as compared to $P(\lambda)$ and $\exp(iE_n(\lambda)\tau_{\lambda}/\lambda)$ so that they can be approximated by the zeroth order terms of the corresponding Taylor expansions around $\lambda=\hbar$ and taken outside the integral, and denoted ${\rho_Q}_{mn}(0)\doteq\langle\phi_m(\hbar)|\hat{\rho}(0)|\phi_n(\hbar)\rangle$. Such an approximation is valid when the width of the fluctuation of $\lambda$ around $\hbar$ is sufficiently small. For the model with $P(\lambda)$ given by Eq. (\ref{log-normal pdf random variable}), it is attained by assuming a sufficiently small $\sigma^2$. 

One thus needs to evaluate the following type of integral 
\begin{equation}
D_{mn}(\tau_{\lambda})\doteq\int_{0}^{\infty} d\lambda P(\lambda)e^{-\frac{i}{\lambda}(E_m(\lambda)-E_n(\lambda))\tau_{\lambda}},  
\label{decay factor 1}
\end{equation}
where, since the integrand is an even function of $\lambda$, then it is sufficient to evaluate the integral along the positive axis of $\lambda$. Let us then expand $E(\lambda)/\lambda$ around $\lambda=\hbar$ up to the first order to have 
\begin{eqnarray}
\frac{E_n(\lambda)}{\lambda}\approx \frac{E_n(\hbar)}{\hbar}\Big(2-\lambda/\hbar\Big),
\label{l'proximasione}
\end{eqnarray}
where we have approximated $E_n(\lambda)$ by its zeroth order term $E_n(\hbar)$. Inserting Eq. (\ref{l'proximasione}), Eq. (\ref{decay factor 1}) can then be approximated as 
\begin{equation}
D_{mn}(\tau_{\lambda})\approx\int_0^{\infty} d\lambda P(\lambda)e^{-\frac{i}{\hbar}E_{mn}(\hbar)\tau_{\lambda}(2-\lambda/\hbar)}, 
\label{decay factor 2} 
\end{equation}  
where $E_{mn}(\hbar)\doteq E_m(\hbar)-E_n(\hbar)$ is just the quantum mechanical energy difference.  

To further evaluate Eq. (\ref{decay factor 2}), let us first make a coordinate transformation $x=\ln\lambda$. Then Eq. (\ref{decay factor 2}) becomes  
\begin{eqnarray}
D_{mn}(\tau_{\lambda})\approx\int_{-\infty}^{\infty}dxf(x)\exp\Big\{-\frac{i}{\hbar}E_{mn}(\hbar)\tau_{\lambda}(2-e^{x-\mu})\Big\},
\label{suppression integral 2}
\end{eqnarray}
where $f(x)=e^xP(e^x)$. Recalling again that the width of the fluctuation of $\lambda$ around $\hbar$ is assumed to be sufficiently small, then only $\lambda$ with  $\lambda/\hbar=e^{x-\mu}\approx 1$ gives a non-negligible contribution to the integral of Eq. (\ref{suppression integral 2}). In this case, one has $|x-\mu|\ll 1$ so that one can expand up to the first order $2-e^{x-\mu}\approx 1-(x-\mu)$. Inserting into Eq. (\ref{suppression integral 2}), one thus has  
\begin{eqnarray}
D_{mn}(\tau_{\lambda})\approx e^{-\frac{i}{\hbar}E_{mn}(\hbar)\tau_{\lambda}}D_{mn}^f(E_{mn}(\hbar)\tau_{\lambda}/\hbar),
\label{suppression integral 3}
\end{eqnarray}
where $D_{mn}^f(E_{mn}(\hbar)\tau_{\lambda}/\hbar)$ is the Fourier transform of $f$ defined as 
\begin{equation}
D_{mn}^f(E_{mn}(\hbar)\tau_{\lambda}/\hbar)\doteq\int_{-\infty}^{\infty}dxf(x)\exp\Big\{\frac{i}{\hbar}E_{mn}(\hbar)\tau_{\lambda}(x-\mu)\Big\}. 
\label{decaying Fourier integral}
\end{equation}  

Let us apply Eqs. (\ref{suppression integral 3}) and (\ref{decaying Fourier integral}) to the above log-normal statistical model. In this case, $P(\lambda)$ in Eq. (\ref{decay factor 2}) is given by the log-normal distribution of Eq. (\ref{log-normal pdf random variable}) so that $f(x)$ in Eq. (\ref{decaying Fourier integral}) takes the form of a Gaussian $f(x)=(2\pi\sigma^2)^{-1/2}\exp(-(x-\mu)^2/2\sigma^2)$. Evaluating Eq. (\ref{decaying Fourier integral}), one thus obtains the following Gaussian decay
\begin{equation} 
D_{mn}^f(E_{mn}(\hbar)\tau_{\lambda}/\hbar)=\exp\Big\{-\frac{\sigma^2}{2}\Big(\frac{E_{mn}(\hbar)}{\hbar}\Big)^2\tau_{\lambda}^2\Big\}, 
\label{Gaussian decay}
\end{equation}
which occurs for an interval of time $\tau_{\lambda}$. Putting Eqs. (\ref{Gaussian decay}) and (\ref{suppression integral 3}) into Eq. (\ref{first integration 1}) one has
\begin{eqnarray}
\hat{I}_1\approx\sum_{mn}|{\phi_m}(\hbar)\rangle{\rho_Q}_{mn}(0)\langle{\phi_n}(\hbar)|\nonumber\\
\times e^{-\frac{i}{\hbar}E_{mn}(\hbar)\tau_{\lambda}}e^{-\frac{\sigma^2}{2}\big(\frac{E_{mn}(\hbar)}{\hbar}\big)^2\tau_{\lambda}^2}.  
\label{first integration 2}
\end{eqnarray}  
 
In the same way, we can then carry out the integration over $\lambda_2$ in Eq. (\ref{path integral for density matrix}) 
\begin{eqnarray} 
\hat{I}_2\doteq \int d\lambda_2P(\lambda_2)\hat{U}_{\lambda_2}(\tau_{\lambda})\hat{I}_1\hat{U}_{\lambda_2}^{\dagger}(\tau_{\lambda})\hspace{30mm}\nonumber\\
=\sum_{jk}\int d\lambda_2P(\lambda_2)\hat{U}_{\lambda_2}(\tau_{\lambda})|\phi_j(\lambda_2)\rangle\langle\phi_j(\lambda_2)|\hat{I}_1\hspace{10mm}\nonumber\\
\circ|\phi_k(\lambda_2)\rangle\langle\phi_k(\lambda_2)|\hat{U}_{\lambda_2}^{\dagger}(\tau_{\lambda}). 
\label{second integration} 
\end{eqnarray}  
Performing the same approximation as before, one has 
\begin{eqnarray}
\hat{I}_2\approx\sum_{jk}|\phi_j(\hbar)\rangle\langle\phi_j(\hbar)|\hat{I}_1|\phi_k(\hbar)\rangle\langle\phi_k(\hbar)|\nonumber\\
\times e^{-\frac{i}{\hbar}E_{jk}(\hbar)\tau_{\lambda}}e^{-\frac{\sigma^2}{2}\big(\frac{E_{jk}(\hbar)}{\hbar}\big)^2\tau_{\lambda}^2}. 
\label{second integration 1}
\end{eqnarray} 
Inserting Eq. (\ref{first integration 2}), one gets
\begin{eqnarray}
\hat{I}_2\approx\sum_{jk}|\phi_j(\hbar)\rangle{\rho_Q}_{jk}(0)\langle\phi_k(\hbar)|\nonumber\\
\times e^{-\frac{i}{\hbar}E_{jk}(\hbar)2\tau_{\lambda}}e^{-\frac{\sigma^2}{2}\big(\frac{E_{jk}(\hbar)}{\hbar}\big)^22\tau_{\lambda}^2}. 
\label{second integration 2}
\end{eqnarray} 

This result can then be used to perform the next integration over $\lambda_3$ and so on. After $n$ times of integrations, one finally has 
\begin{eqnarray}
\widetilde{\rho}(t)\approx\sum_{rs}|\phi_r(\hbar)\rangle{\rho_Q}_{rs}(0)\langle\phi_s(\hbar)|\nonumber\\
\times e^{-\frac{i}{\hbar}E_{rs}(\hbar)n\tau_{\lambda}}e^{-\frac{\sigma^2}{2}\big(\frac{E_{rs}(\hbar)}{\hbar}\big)^2n\tau_{\lambda}^2}\nonumber\\
=\sum_{rs}|\phi_r(\hbar)\rangle{\rho_Q}_{rs}(0)\langle\phi_s(\hbar)|\nonumber\\
\times e^{-\frac{i}{\hbar}E_{rs}(\hbar)t}e^{-\frac{\sigma^2}{2}\big(\frac{E_{rs}(\hbar)}{\hbar}\big)^2\tau_{\lambda} t},
\label{n integration}
\end{eqnarray}
where in the second line we have used $t=n\tau_{\lambda}$. The $rs-$element of the $\lambda-$averaged density matrix in energy basis at time $t$ is thus given by 
\begin{eqnarray}
\widetilde{\rho}_{rs}(t)\doteq\langle\phi_r(\hbar)|\widetilde{\rho}|\phi_s(\hbar)\rangle\approx{\rho_Q}_{rs}(t)e^{-\frac{\sigma^2}{2}\big(\frac{E_{rs}(\hbar)}{\hbar}\big)^2\tau_{\lambda} t},
\label{decaying density matrix}
\end{eqnarray}
where ${\rho_Q}_{rs}(t)={\rho_Q}_{rs}(0)e^{-\frac{i}{\hbar}E_{rs}(\hbar)t}$ is the $rs-$element of the quantum mechanical density matrix in energy basis at time $t$. Hence there is a decaying factor in time: the off-diagonal elements of the $\lambda-$averaged density matrix in energy basis is suppressed in time and the suppression is stronger for elements with larger energy difference, and the diagonal elements are given by those of the quantum mechanical density matrix. There is thus in general an `intrinsic' or `fundamental' decoherence even if the system is quantum mechanically closed. It gives an additional loss of coherence to the well-known environment-induced-decoherence \cite{Zeh 1,Zurek 1,Joos book,Schlosshauer,Nieuwenhuizen}. 
  
There are three parameters that determine the rate of loss of coherence in Eq. (\ref{decaying density matrix}): $\hbar$, $\sigma$ and $\tau_{\lambda}$. All characterize the fluctuation of $|\lambda|$ and are undetermined in the model. $\tau_{\lambda}$ is the time scale of the fluctuation of $|\lambda|$, while $\hbar$ and $\sigma$ determine its average and mean deviation. To fit experimental results, $\hbar$ has to be numerically identified by the reduced Planck constant. Fixing $\tau_{\lambda}$, one can see that the exponential decay is stronger for larger value of $\sigma$, and in the limit of vanishing $\sigma$, one has an absolute quantum coherence regardless the value of $\tau_{\lambda}$ and the scale of the energy difference. 

To determine the numerical values of $\sigma$ and $\tau_{\lambda}$, one may resort to experiments or by developing a deeper theory which unveils the physical origin of the chaotic fluctuation of infinitesimal stationary action in the statistical model. To this end, recall that, as discussed at the end of subsection \ref{global lambda}, $\lambda$ is left unfixed by the principle of Locality so that it is a global variable spatially uniform across the universe. One may thus argue that its statistics might be related to the global property of the latter. If this is indeed the case, then $\hbar$, $\sigma$ and $\tau_{\lambda}$ should be determined by first devising a deeper level theory based on a proper cosmological theory as the starting point to develop the deterministic model of quantum fluctuation. 

As mentioned before, an argument of this kind, that quantization is cosmological in origin, is advanced by Calogero in Ref. \cite{Calogero} within the framework of a general class of stochastic models of quantization, in which quantum fluctuation is argued to be caused by a universal gravitational fluctuation. The latter in turn arises due to the universality of long range gravitational interaction, the assumption of the granularity of universe (its main components are particles), and the universal chaoticity of the classical many body systems. He then derived a formula which expresses the Planck constant in terms of cosmological quantities and gravitational constant which remarkably yields the correct-order-magnitude for the Planck constant. He has also suggested that a random fluctuation around $\hbar$ is viable and may give the origin of ultra-weak violation of time-reversal symmetry. Following Calogero, one may therefore argue that the chaotic fluctuation of infinitesimal stationary action in the statistical model discussed in the present paper is due to a universal gravitational fluctuation. In this sense, the statistical model may also be seen to give a justification for Calogero's program.   

In view of the above observation, let us write the elements of the density matrix of Eq. (\ref{decaying density matrix}) as  
\begin{equation}
\widetilde{\rho}_{rs}(t)\approx{\rho_Q}_{rs}(t)\exp\Big\{-\frac{\sigma^2}{2}\Big(\frac{\omega_{rs}}{\omega_{\lambda}}\Big)^2\frac{t}{\tau_{\lambda}}\Big\},
\label{universal decay}
\end{equation}
where $\omega_{rs}=E_{rs}(\hbar)/\hbar$ is a frequency associated to energy difference $E_{rs}(\hbar)$ characterizing the system under interest and $\omega_{\lambda}=1/\tau_{\lambda}$ is the effective frequency of the fluctuation of $|\lambda|$. Since $\tau_{\lambda}$ and its inverse $\omega_{\lambda}$ are argued to be determined by the global structure of the universe, then the strength of the decay is determined by the relative scale of the system under interest with that of the universe. In the case of microscopic systems, for example, it is reasonable to assume that $|\omega_{rs}/\omega_{\lambda}|$ is very small so that the coherence is kept for a long time thus observable in experiment. By contrast, for macroscopic system, one may argue that $|\omega_{rs}/\omega_{\lambda}|$ is sufficiently large leading to a fast destruction of coherence.    

One can finally show by substitution that Eq. (\ref{n integration}) satisfies the following master equation:
\begin{equation}
\partial_t\widetilde{\rho}=-\frac{i}{\hbar}[\hat{H}_{\hbar},\widetilde{\rho}]-\alpha^2[\hat{H}_{\hbar},[\hat{H}_{\hbar},\widetilde{\rho}]], 
\label{Lindblad equation energy-conserving}
\end{equation}
where $\alpha=(\sigma/\hbar)\sqrt{\tau_{\lambda}/2}$. It belongs to a class of Lindblad equation \cite{Lindblad,Sudarshan derivation} where the Lindblad operator is linearly proportional to the quantum Hamiltonian: $\hat{L}=\alpha\hat{H}_{\hbar}$. In the limit of vanishing $\alpha$, the second decoherence term of the right hand side is vanishing and one regains the von-Neumann equation describing a unitary time evolution. It is easy to show that for $\widetilde{\rho}$ satisfying the above Lindblad equation, quantum mechanically conserved quantities are kept conserved. In particular, the average energy identified as $\mbox{Tr}\{\hat{H}_{\hbar}\widetilde{\rho}\}$ is a constant of motion. Moreover, it is well-known that for such $\tilde{\rho}$, the Shannon entropy defined as $-\mbox{Tr}\{\widetilde{\rho}\ln\widetilde{\rho}\}$ is monotonically increasing with time. 

\section{Concluding remarks} 

We have first discussed a deterministic model of the universal random behavior observed in microscopic world based on a chaotic fluctuation of deviation from infinitesimal stationary action whose distribution is singled out uniquely by the principle of Locality to have an exponential law, up to a parameter $\lambda$ which determines its average as $|\lambda|/2$. The dynamics and statistics of the ensemble of trajectories are then shown to be governed by a generalized Schr\"odinger equation in which the reduced Planck constant $\hbar$ of the usual Sch\"odinger equation is replaced by a chaotically fluctuating parameter $|\lambda|$. The quantum coherence that is governed by the unitary Schr\"odinger equation with Born's statistical interpretation of wave function is thus recovered as a specific case when $|\lambda|=\hbar$ for all the time so that the model is stationary in time and the average deviation from infinitesimal stationary action is given by $\hbar/2$. 

We then proceeded to go beyond the stationary case by allowing $|\lambda|$ fluctuates randomly around $\hbar$ with a finite small width and a finite time scale much smaller than the time scale of observation. We showed that averaging over the distribution of $|\lambda|$ leads to an effective non-unitary time evolution, providing an intrinsic mechanism of decoherence in energy basis, thus keeping the average energy conserved, adding to the usual environmental decoherence. The rate of the decoherence depends on the average, mean deviation and time scale of the fluctuation of $|\lambda|$. Coherence and decoherence are therefore explained in a unified way as two features of the same statistical model corresponding to microscopic and macroscopic regimes, respectively, necessitating no interaction with external structures. Furthermore, showing that within the statistical model measurement-interaction can be treated in equal footing as the other types of interaction \cite{AgungSMQ7,AgungSMQ9}, we have argued that the objective locality of the model implies no-signaling in the context of measurement. We have thus a class of no-signaling modifications of quantum dynamics which provides an internal mechanism of decoherence in macroscopic regime.  We have also suggested that the statistical fluctuation of $|\lambda|$ may have a cosmological origin. 

The above results indicate that the currently recommended value of Planck `constant' might be {\it not} exact, but is an extremely accurate extrapolation of a chaotically fluctuating parameter $|\lambda|$ of the present statistical model. They also open a possibility that the numerical value of $\hbar$ itself may fluctuate with (cosmic) time. A similar suggestion is also offered by Calogero in Ref. \cite{Calogero}. It is then imperative to develop experimentally accessible physical systems which can probe such possible fluctuation, and which can give testable predictions that are different from those of the existing models of intrinsic decoherence \cite{Milburn,Garay,Gambini 1,Ellis 1,Diosi decoherence,Gambini QG}. 

It is also very interesting to compare the statistical model with the stochastic dynamical reduction models (SDRM), say the one reported in Ref. \cite{Pearle}. First, in the statistical model, we have {\it derived} a chaotically parameterized Schr\"odinger equation: $d\Psi=-(i/|\lambda|)\hat{H}_{|\lambda|}\Psi dt$. In the SDRM, on the other hand, one {\it postulates} an additional stochastic term to the original Schr\"odinger equation to have $d\Psi_Q=-(i/\hbar)\hat{H}_{\hbar}\Psi_Qdt+f(dW)\Psi_Q$, where $dW$ is an increment of Wiener process. Let us also note that while our modification is naturally suggested by the reconstruction of quantum mechanics, the modification in SDRM is rather ad-hoc. The main motivation of the development of the SDRM is to have an intrinsic mechanism of dynamical wave function collapse. By contrast, the statistical model is a no-collapse hidden variable model. Both models also show a similar behavior of decoherence in macroscopic regime. It is therefore interesting to ask if one can further modify the statistical model by introducing some Wiener processes to arrive at the equation of SDRM. 

\begin{acknowledgments} 

The author acknowledge constructive comments from an anonymous referee. 

\end{acknowledgments}

\end{document}